\documentclass[12pt,preprint]{aastex}

\newcommand{\tlusty}{{\sc Tlusty}}
\newcommand{\synspec}{{\sc Synspec}}
\newcommand{\teff}{$T_{\rm eff}$}
\newcommand{\grid}{{\sc Ostar2002}}
\newcommand{\Mdot}{\hbox{$\dot M$}}
\newcommand{\kms}{km\,s$^{-1}$}

\newcommand{\lb}{${\lambda}$}
\newcommand{\msolyr}{M$_{\odot}$\,yr$^{-1}$}

\newcommand{\vsini}{$V\sin i$}

\newcommand{\vinf}{$v_{\infty}$}

\newcommand{\halpha}{H$\alpha$}

\newcommand{\evi}{$\times\,10^{-6}$}
\newcommand{\evii}{$\times\,10^{-7}$}
\newcommand{\eviii}{$\times\,10^{-8}$}
\newcommand{\eix}{$\times\,10^{-9}$}

\slugcomment{\today}

\shorttitle{Quantitative Spectroscopy of O~Stars at Low Metallicity.}
\shortauthors{Bouret et al.}

\begin{document}

\title{Quantitative Spectroscopy of O~Stars at Low Metallicity. \\
       O Dwarfs in NGC 346.\altaffilmark{1}}
          
\altaffiltext{1}{Based on observations with the NASA/ESA Hubble Space
      Telescope, obtained at the Space Telescope Science Institute, which
      is operated by the Association of Universities for Research in
      Astronomy, Inc., under NASA contract NAS5-2655. Also based on
      observations obtained at the European Southern Observatory
      (La Silla) and at the Anglo-Australian Observatory (Siding Spring).}
          
\author{J.-C. Bouret\altaffilmark{2}, T. Lanz\altaffilmark{3,4},
         D. J. Hillier\altaffilmark{5}, S. R. Heap\altaffilmark{3},
         I. Hubeny\altaffilmark{3,6}, D. J. Lennon\altaffilmark{7},
         L. J. Smith\altaffilmark{8}, and C. J. Evans\altaffilmark{7,8}}

\altaffiltext{2}{Laboratoire d'Astrophysique de Marseille,
       Traverse du Siphon - BP 8, 13376 Marseille Cedex 12, France}
\altaffiltext{3}{NASA Goddard Space Flight Center, Code 681, Greenbelt, MD 20771}
\altaffiltext{4}{Department of Astronomy, University of Maryland, College Park, MD 20742}
\altaffiltext{5}{Department of Physics and Astronomy, University of Pittsburgh,
       Pittsburgh, PA 15260}
\altaffiltext{6}{National Optical Astronomy Observatories, Tucson, AZ 85726}
\altaffiltext{7}{Isaac Newton Group, Apartado 321, 38700 Santa Cruz de La Palma,
       Canary Islands, Spain}
\altaffiltext{8}{Department of Physics and Astronomy, University College London,
       Gower Street, London WC1E 6BT, UK}

\email{Jean-Claude.Bouret@astrsp-mrs.fr}
\email{lanz@stars.gsfc.nasa.gov}

\begin{abstract}
We present the results of a detailed analysis of the
properties of dwarf O-type stars in a metal-poor environment.
High-resolution, high-quality, ultraviolet and optical spectra of six
O-type stars in the \ion{H}{2} region NGC 346 have been obtained from
a spectroscopic survey of O stars in the SMC. Stellar parameters and
chemical abundances have been determined using NLTE line-blanketed
photospheric models calculated with \tlusty. Additionally, we have modeled
the spectra with the NLTE line-blanketed wind code, CMFGEN, to derive
wind parameters. Stellar parameters and chemical abundances, and in
particular iron abundances, obtained with the two NLTE codes compare
quite favorably. This consistency demonstrates that basic photospheric parameters
of main-sequence O stars can be reliably determined using NLTE static
model atmospheres. With the two NLTE codes, we need to introduce a
microturbulent velocity in order to match the observed spectra. Our
results hint at a decrease of the required microturbulent velocity
from a value close to the sonic velocity in early O stars to a low
value in late O stars. Similarly to several recent studies of Galactic,
LMC and SMC stars, we derive effective temperatures lower than predicted
from the widely-used relation between spectral type and \teff, resulting
in lower stellar luminosities and lower ionizing fluxes. From evolutionary
tracks in the HR diagram, we find an age of 3\,$10^6$ years for NGC~346.
A majority of the stars in our sample reveal CNO-cycle processed material
at their surface during the main-sequence stage, indicating thus fast
stellar rotation and/or very efficient mixing processes. We obtain an overall
metallicity, $Z = 0.2 Z_\odot$, in good agreement with other recent analyses
of SMC stars. We study the dependence of the mass loss rate with the stellar
metallicity and find a satisfactory agreement with recent theoretical predictions
for three most luminous stars of the sample. The wind-momentum luminosity
relation for our sample stars derived for these stars agree with previous studies.
However, the three other stars of our sample reveal very weak signatures of mass loss.
We obtain mass loss rates that are significantly lower than 10$^{-8}\,$\msolyr,
which is below the predictions of radiative line-driven wind theory by an order of
magnitude or more. Furthermore, evidence of clumping in the wind of main-sequence of
O stars is provided by \ion{O}{5}\,\lb1371. Like previous studies of O star winds,
we are unable to reproduce this line with homogeneous wind models, but we have achieved
very good fits with clumped models. Clumped wind models systematically yield lower
mass loss rates than theoretical predictions.
\end{abstract}


% -----------------------------------------------------------------------------

\section{Introduction}
\label{IntroSect}
   
Despite their rarity, hot massive stars are prominent contributors
to the dynamics and energetics of the ISM and to the global evolution
of their host galaxies. The optical spectra of distant, 
star-forming galaxies exhibit numerous features typical
of the UV spectra of O and B stars \citep{conti96, steidel96}.
In order to determine the
properties of starbursts galaxies at high redshift, it is thus
essential to understand the physical properties and evolution
of massive stars in a low metallicity environment. O stars in the
Small Magellanic Cloud offer the best opportunity to achieve this goal.
Indeed, the SMC is a galaxy which is little evolved having a relatively
low metal content,
and is sufficiently close to obtain high-quality UV and optical spectra of
individual stars. Additionally, modeling tools to analyze the photosphere and
winds of hot, massive stars with a high level of accuracy and reliability have
become available in recent years. Major progress has been achieved to model
the stellar photosphere and stellar wind in an unified approach, as well as
to incorporate a sophisticated treatment of NLTE line blanketing that
accounts for all the important opacity sources in the stellar atmospheres
(e.g., programs CMFGEN, Hillier~\& Miller 1998; WM-BASIC,
Pauldrach et al. 2001; \tlusty, Hubeny \& Lanz 1995). 

Winds from massive stars are radiatively driven through the transfer of
momentum from the stellar radiation field to the atmospheric material via
photon absorptions and scattering due to thousands of metal lines
\citep{cak75, kudr00}.
An immediate consequence of the nature of these winds is that the
wind properties (mass loss rate, terminal velocity) depend on the
stellar metallicity. For example, \citet{vink01} have predicted that 
the mass loss rate scales as \Mdot\ $\propto Z^{0.85}$, for metallicities
ranging from 1/30 to 3 times the solar metallicity. The agreement between
their predicted values and the measured mass loss rates for samples of
O-type stars in the Milky Way is good \citep{vink00}, but
they do not agree well for LMC stars and a large scatter is found
for SMC stars \citep{vink01}.
In this context, it is of special interest to derive accurate
estimates of both elemental abundances and mass loss rates from the
analysis of individual O stars in the SMC, in order to provide an important
scaling law for evolutionary models of massive stars. Additionally,
such analyses yield the quantities to further test the Wind-momentum Luminosity
Relation \citep{puls96} in a low metallicity environment.

Following the pioneering studies of SMC OB stars in the UV based
on low resolution and limited signal-to-noise ratio data recorded by
{\sl IUE\/} \citep{smith97} or HST/FOS \citep{walborn95, puls96, haser98},
Lennon led a spectroscopic survey of a score of SMC OB stars at higher
spectral resolution in the UV with HST/STIS \citep{walborn00}.
About half of the stars are located in NGC~346, the largest \ion{H}{2} region
in the SMC which has been extensively observed with ground-based telescopes
in the past \citep{walborn78, walborn86, niemela86, massey89}. These
new observations in the UV have been supplemented by optical spectroscopy.

Taking advantage of these new, high-quality spectroscopic data,
\citet{AV83} focussed their interest on describing the wind properties of
O stars. They performed a detailed study of two stars, AV~83 and AV~69,
which have similar effective temperatures, luminosities and metallicities,
but show very different wind signatures. Their analysis indicates that
the O7\,Iaf+ star AV~83 has a slow, dense wind, which most likely is highly clumped.
Moreover, AV~83 reveals a substantial enhancement of its nitrogen surface
abundance, consistent with the presence at the surface of material processed
internally by the CNO cycle. On the other hand, the OC7.5\,III((f)) star
AV~69 has a substantially less dense wind and has a surface composition
similar to SMC gas. These results are consistent with our current understanding
of stellar evolution (e.g. Maeder~\& Meynet 2000) if AV~83
is a fast rotator which has experienced rotationally enhanced mass loss
and rotationally induced mixing. On the other hand, AV~69 would be a slow
rotator. \citet{AV83} have systematically examined the
influence of various parameters on the predicted spectrum, providing thus
a solid basis for further studies of O star winds.

We present in this paper the result of our spectroscopic analysis, from the
far-UV to the visible, of six O dwarf stars in NGC~346.
We describe the observations and data reduction in Sect.~\ref{ObsSect}.
Section~\ref{AssumSect} discusses the model atmospheres used in our analysis,
while the methodology used to derive the stellar parameters, chemical abundances,
and winds parameters is exposed in \S\,\ref{MethodSect}. We discuss our results
and related uncertainties, star by star in Sect.~\ref{ResuSect}. We put then
our results in a broader astrophysical context in Sect.~\ref{DiscSect},
comparing in particular the derived mass loss rates to theoretical predictions
and discussing their dependence with the stellar metallicity. The paper
closes on general conclusions (\S\,\ref{ConclSect}) based on this analysis of
O dwarfs in the SMC low metallicity environment.

% ------------------------------------------------------------------------------

\section{Observational Background}
\label{ObsSect}
   
High-resolution, high-quality UV and optical spectra of a score of O stars in
the SMC have been recorded to serve as a spectral template of a young stellar population
in a low metallicity environment. A full description of the whole dataset and
data reduction can be found in \citet{walborn00}. The sample
was selected with a bias towards stars with a sharp-lined spectrum, thus towards
stars viewed preferentially pole-on or possibly towards intrinsically slow rotators.
In this paper, we are concerned with six stars in NGC~346 that are on
the main sequence.
We only recall here the major characteristics and some relevant information about
the selected stars. Table~\ref{obsstis} summarizes the observational parameters of
the stellar sample. Stars' identification (MPG number) is from \citet{massey89};
spectral types and the photometry are taken from \citet{walborn00}.
Following \citet{walborn02}, the hottest star of our sample, NGC~346 MPG 355, belongs
to the newly-defined O2 spectral class and has been classified O2~III\,(f*). Formally,
it is not an O dwarf, but dwarf and giant stars occupy nearly the
same location in the HR diagram at the hot end of the main sequence. The inclusion
of this star in our sample allows us to sample a range from the earliest to the latest
O stars. The spectral type of NGC~346 MPG 487 was attributed by Walborn (priv. comm.) based
on the new visual spectroscopy discussed below. Radial velocities have been measured by shifting
the observed spectra relative to the model spectra. Our radial velocities are very close to
the values listed by \citet{walborn00}.

\subsection{UV Spectroscopy}
\label{UVSpecSect}

Ultraviolet spectra have been obtained in the framework of {\sl HST\/} General Observer
program 7437 (PI: D. J. Lennon) for 19 O-type stars in the SMC. All program stars
have been observed with STIS and the far-UV MAMA detector in the E140M mode, through the
0\farcs 2$\times$0\farcs 2 aperture. A spectral interval from 1150 to 1700\,\AA\ 
is covered in a single exposure, at an effective resolving power, $R=46\,000$.
The STIS data have been reduced at Goddard Space Flight Center, using the CALSTIS
software developed by the STIS Investigation Definition Team \citep{calstis}.
\citet{walborn00} have described the reduction process in detail. The spectra
are then normalized, because the predicted flux distribution in the far-UV not only
depends on the assumed SMC distance but is very sensitive to the adopted
correction of interstellar extinction.

\subsection{Optical Spectroscopy}
\label{VisSpecSect}

Complementary optical spectroscopy was obtained for all program stars
\citep{walborn00}. Spectra of four stars of this sample (\objectname{NGC 346 MPG 355},
\objectname{NGC 346 MPG 324}, \objectname{NGC 346 MPG 368}, \objectname{NGC 346 MPG 487})
have been obtained at Anglo-Australian Telescope (AAT) with the echelle
spectrograph UCLES. The spectral interval ranges from 3847 to 5008\,\AA,
at a typical resolution, $R=25\,000$. The seeing varied between 0\farcs 8 and 2\arcsec.
Additional observations of the red spectrum have been obtained to constrain the
mass loss rate with H$\alpha$ using the same instrumentation at a similar resolution.
Observations of H$\alpha$ have been secured for the three stars of our sample that
show marked UV spectral signatures of a stellar wind (MPG 355, 324, 368). 

The two other stars, \objectname{NGC 346 MPG 113}, and \objectname{NGC 346 MPG 12},
were observed with ESO 3.6\,m telescope equipped with CASPEC. The spectra cover an
interval from 3910 to 5170\,\AA, at a similar spectral resolution, $R=25\,000$.

The echelle spectral data were reduced uniformly with the FIGARO package
by the UCL group. Details on the procedure can be found in \citet{walborn00}.

% ------------------------------------------------------------------------------

\section{Models and Assumptions}
\label{AssumSect}

Strong stellar winds, significant departures from the Local
Thermodynamic Equilibrium (LTE) and an effect of numerous
metal lines, traditionally called metal line blanketing,
present the major difficulties in modeling the atmospheres of
O-type stars. These three issues must be addressed by
the model atmospheres in order to provide reliable photospheric
and wind parameters of massive stars. This involves constructing
fully-blanketed, NLTE models for an entire stellar atmosphere,
going from a quasi-static photosphere out to the supersonic wind.
Considerable advances in modeling techniques have been achieved
during recent years (for a number of recent reviews, see Hubeny
et~al. 2003), and several computer
programs that can handle this problem are available now --
program CMFGEN \citep{CMFGEN}; PHOENIX \citep{PHOE97};
Munich programs \citep{Mun01}; and the Kiel-Potsdam code \citep{LK02}.

Unified models, with a consistent treatment of the photosphere and
the wind, are obviously necessary to analyze the P~Cygni profile
of strong lines in the UV spectra of O stars and then derive the
basic wind parameters (mass loss rate, terminal velocity). However,
hydrostatic, fully-blanketed, NLTE photospheric models may in many cases
remain a preferable alternative to unified models with a simplified
treatment of a photosphere. Indeed, the bulk of spectral lines in
O stars are formed in the photosphere where velocities are small and
geometrical extension is negligible. The spectrum of slowly-rotating
O stars show mostly narrow and symmetric lines (see, e.g., NGC~346 MPG~113,
Fig.~\ref{P113Fig}). This statement is valid in all
but the most extreme cases of stellar winds, like those observed in Wolf-Rayet
stars and extreme Of supergiants. Moreover, winds are weaker in low-metallicity
environments making photospheric models even more relevant in this case.
Additionally, there are many remaining uncertainties as to the exact properties of
stellar winds of hot stars, even though the paradigm of radiatively-driven
winds is well established. For example, we do not have in many cases a consistent
solution: the calculated line force is often too small to drive a wind through
the critical point (though
some progress is being made), so a $\beta$-type velocity law must be adopted
to describe empirically the velocity and density structure of the wind. While
mathematically trivial, it is physically not so obvious how to connect the photospheric
exponential density structure with the wind power law (see Hillier et al. 2003
for a discussion); some spectral features are formed in the connection region. Furthermore,
radiative equilibrium is not really satisfied in these winds, where shocks dump
mechanical energy and heat the wind, thus changing its ionization structure
(e.g., super-ionization). Finally, the assumption of one-dimensional geometry is
challenged by the likely presence of dense clumps in the wind and by rotation;
the role of magnetic fields is not yet understood, and their importance in
structuring the wind will very likely be more important than in the photosphere.

Therefore, we have performed an analysis using model atmospheres calculated
with the photospheric program, \tlusty, and the unified model code, CMFGEN. We intend
to compare the parameters derived from these two NLTE model atmosphere programs;
we may consider them as two independent programs although the atomic data used
by the two codes come mostly from the same sources. Second, we will examine the
effect of an extended stellar wind on the photospheric lines in O dwarf spectra.
We describe thereafter the basic characteristics of these two codes.

\subsection{Photospheric Models}
\label{TlustySect}

Photospheric models have been constructed with the model atmosphere
code, \tlusty\  \citep{NLTE1}. The main assumptions
of the code are a plane-parallel geometry, hydrostatic equilibrium,
and radiative equilibrium. Departures from LTE are explicitly
allowed for a large set of chemical species and arbitrarily complex
model atoms. Line opacity is treated in detail using an Opacity Sampling
technique with close to 200\,000 frequency points over the whole spectrum.
For this work, we have extracted models from the new, extensive \grid\ 
grid \citep{OS02}. These fully-blanketed, NLTE model atmospheres
include about 100\,000 individual atomic levels from 45 ions (\ion{H}{1-ii},
\ion{He}{1-iii}, \ion{C}{2-v}, \ion{N}{2-vi}, \ion{O}{2-vii}, \ion{Ne}{2-v},
\ion{Si}{3-v}, \ion{P}{4-vi}, \ion{S}{3-vii}, \ion{Fe}{3-vii}, \ion{Ni}{3-vii}),
as summarized in Table~\ref{atom_mod}.
We have selected models with a scaled-solar composition, $Z/Z_\odot =$ 1/5, and 1/10,
that are appropriate for SMC stars. We have adopted the solar abundances from
\citet{Sun98}. All model atmospheres assume an helium abundance,
$y$ = He/H = 0.1 by number\footnote{Throughout this paper, chemical abundances are quoted
by number density relative to hydrogen.}, and a microturbulent velocity,
$\xi_{\rm t}=10\,$\kms. The \grid\  model atmospheres are described in detail
by \citet{OS02}.

\citet{OS02} have pointed out that the omission of highly-excited atomic
levels of light species in \grid\  model atmospheres results in underestimating
some recombination rates. We have therefore recalculated a limited number of model
atmospheres with more extensive model atoms for \ion{C}{3} and
\ion{N}{4} in order to achieve a better prediction of recombination lines
seen in emission in the visible spectrum.

\tlusty\  solves for the full photospheric structure, providing the temperature
and density stratification and the NLTE populations. A detailed synthetic
spectrum is then calculated with \synspec, varying if necessary the abundance
of individual species (e.g., the N/C abundance ratio), or the microturbulent
velocity. However, we have always imposed the condition
that the photospheric turbulent velocity remains
smaller than the photospheric speed of sound (typically, $c_{\rm sound} \leq$ 25\,\kms).

\subsection{Wind Models}
\label{CMFGENSect}

Spherically symmetric wind models have been constructed with the
NLTE code CMFGEN \citep{CMFGEN, TuebJDH}. This code
solves the radiative transfer equation in the co-moving frame,
together with statistical equilibrium equations. The wind models
include 28 explicit ions of H, He, C, N, O, Ne, Si, P, S, and Fe.
A total of 3300 individual levels grouped into 900 superlevels
are included, with a full array of over 20\,000 bound-bound transitions.
The ionization stages, number of individual levels and superlevels of
each model atom are summarized in Table~\ref{atom_mod}. We have
adopted an atomic data set that is consistent
with the data set used by \citet{AV83}. Radiative equilibrium has
been assumed to determine the temperature structure throughout the
atmosphere.

CMFGEN does not solve the momentum equation, and a density or velocity
structure is therefore required beforehand. The velocity in the stellar
wind is parameterized with a classical $\beta$-type law which is connected
to an hydrostatic density structure at depth. We have taken the hydrostatic
structure from \tlusty\  models. At the connecting point, we
require that the velocity and the velocity gradient both match.
O star model spectra are relatively sensitive to the exact
density structure which is assumed around the sonic point if the wind
is dense enough. \citet{AV83} have presented a detailed
discussion of this issue for AV~83. Since few lines show evidence of a wind
in our stellar sample, this problem is likely not as severe as in AV83
and most diagnostic lines should be insensitive to it.

We have assumed a depth-independent Doppler profile for all lines when
solving for the atmospheric structure in the co-moving frame.
\citet{martins02} have shown that the resulting
atmospheric structure, level populations, and emergent spectrum, is
not changed by this approximation. In this step, we used the
microturbulent velocity, $\xi_{\rm t}^{\rm phot}$, derived from
\tlusty\  analysis. In the final calculation of the emergent
spectrum in the observer's frame, we have however adopted a
radially-dependent turbulence, expressed as
\begin{equation}
\label{VturbEq}
\xi_{\rm t}(r) = \xi_{\rm t}^{\rm phot} + (\xi_{\rm t}^{\rm max}
-\xi_{\rm t}^{\rm phot})\frac{v(r)}{v_{\infty}},
\end{equation}
where $v(r)$ and $v_{\infty}$ are the velocity law and the terminal
velocity of the wind, and $\xi_{\rm t}^{\rm max}$ is the maximum turbulent
velocity (see \S\,\ref{WindParSect}). Such a turbulent velocity law reflects
the effect of shocks due to wind instabilities. Moreover, we have taken into
account incoherent electron scattering and Stark broadening for \ion{H}{1},
\ion{He}{1}, \ion{He}{2} lines.

CMFGEN can account for clumping in the stellar wind which results from
the unstable nature of radiatively-driven winds. Clumping is described assuming
a volume filling factor, $f$, that there is no interclump medium, and that
the clumps are small compared to the mean free path of photons. The filling
factor is parameterized following an exponential decrease
\begin{equation}
f = f_\infty + (1-f_\infty) \exp(v/v_{\rm cl}),
\end{equation}
where $v_{\rm cl}$ is the velocity at which clumping starts. We have assumed that
clumping starts at $v_{\rm cl}=30$\,\kms, just above the sonic point. The detailed
implementation of clumping in CMFGEN is discussed by \citet{hillier97} and
by \citet{CMFGEN}.

% ------------------------------------------------------------------------------

\section{Methodology}
\label{MethodSect}

We start the analysis using \tlusty\  \grid\  models \citep{OS02}.
The photospheric parameters derived from this analysis provide the initial values
for a second analysis with the unified, wind code, CMFGEN. In this second step, we still
allow for changes in the photospheric parameters, because CMFGEN models also have
a realistic description of photospheric layers. We then compare the consistency
of the two analyses.

\subsection{Determination of Photospheric Parameters}
\label{TeffSect}

Following a long-established way to analyze O star spectra, we first attempt to
match the Balmer \ion{H}{1}, \ion{He}{1} and \ion{He}{2} lines between
3800 and 5000\,\AA\  to determine initial estimates of the photospheric parameters
(\teff, $\log g$, $y$ = H/He). The \ion{He}{1} to \ion{He}{2}
line strength ratios are the characteristic \teff\  indicator in main-sequence
O stars, while $\log g$ is constrained from Balmer line wings. A
simultaneous fit of all optical lines of these ions proves however
often elusive. The task is further complicated by a nebular contamination of
the Balmer and \ion{He}{1} lines since NGC~346 is embedded in strong nebulosity.
Our initial \teff\ diagnostics lines are thus \ion{He}{1}\,\lb4388, 4471, 4713, 4922,
and \ion{He}{2}\,$\lambda$4200, 4541, 4686.
The latter \ion{He}{2} line is often affected by a stellar wind, leading several
authors to disregard \ion{He}{2}\,$\lambda$4686 as a \teff\  indicator
(e.g. Crowther et al. 2002).
However, the stellar wind of SMC O dwarfs is weak enough (see \S\,\ref{ResuSect})
that this line may be helpful. We have not used specifically
\ion{He}{2}\,\lb1640. On one hand, this line does not provide additional photospheric diagnostics.
On the other hand, we have not observed in our stellar sample much influence of the stellar wind
on this line, mostly because the winds are not dense enough.

In addition to the optical spectrum, we also rely on ionization balance of heavy
elements in the UV spectrum. We need however to disentangle their temperature sensitivity
from other parameters, {\it (i)} the microturbulent velocity (e.g \ion{O}{4}\,$\lambda$1338-43,
\ion{S}{5}\,$\lambda$1502, \ion{Fe}{4-vi} lines), {\it (ii)} the mass loss rate
(\ion{O}{4}\,$\lambda$1338-43, \ion{O}{5}\,$\lambda$1371, \ion{C}{3}\,$\lambda$1176),
not to mention {\it (iii)} the adopted chemical abundances. The \grid\  grid allows us
to study trends and to define the best strategy. The safest approach in limiting
the role of other parameters consists in considering line ratios of two (or more) successive
ionization stages of a given chemical element. Therefore, we have mostly used the
\ion{C}{3}\,$\lambda$1176 to \ion{C}{4}\,$\lambda$1169 ratio and iron ions as primary temperature
diagnostics (see also Heap et~al. 2003a). Comparisons between photospheric (\tlusty) and
wind (CMFGEN) models have also helped to estimate the effect of the stellar wind on
temperature sensitive lines; specific results are discussed in \S\,\ref{ResuSect}.
The \teff\  accuracy  is typically of the order of several percent.

Usual spectroscopic diagnostics of $\log g$ are
Balmer line wings, most especially H$\gamma$. High accuracy determinations of 
$\log g$ which are required to derive accurate stellar masses are difficult to achieve.
Indeed, the line wings change relatively little compared to observational limitations
(nebular emission; spectra with moderate signal-to-noise ratios). Moreover, a
strong stellar wind fills the Balmer lines and thus results in underestimating the surface
gravity \citep{SPKGG93, schaerer94}. This later
problem should however remain limited with weaker winds in low metallicity environments.
We have thus always checked our initial estimates derived from \tlusty\  models
with CMFGEN fits to Balmer lines. We stress finally that the {\em effective gravity}
of the models is significantly lower than the real gravity because of the influence
of the radiation pressure (see Fig.~3 in Lanz \& Hubeny 2003b). The influence of
centrifugal acceleration is however ignored, and hence
the actual gravity is underestimated, particularly for fast stellar rotation.

A simultaneous fit of all \ion{He}{1} and \ion{He}{2} lines is almost never achieved
(see, e.g., Herrero et al. 1992; Hillier et al. 2003), leaving significant
uncertainties on helium abundance determinations. Earlier studies systematically
found enhanced helium abundances \citep{herrero92}, while standard stellar
evolution models (without rotation) do not predict such a surface enrichment
for main-sequence O stars \citep{maeder87a}. \citet{smith98} showed that consistent fits
of the helium lines in O supergiants can be achieved when introducing a microturbulent
velocity, helping to solve the ``helium discrepancy'' problem \citep{herrero92}.
In this study, we have adopted an helium abundance, $y=0.1$, as in \tlusty\  \grid\ 
grid, and we start the analysis assuming a microturbulent velocity, $\xi_{\rm t}$ = 10\,\kms.
There is no indication of helium enrichment in our stellar sample.

\subsection{Stellar Luminosities and Radii}
\label{LuminSect}

Absolute visual magnitudes were calculated from the photometry in Table~\ref{obsstis},
assuming a distance modulus of 19.1\,mag (i.e., a distance of 66.1\,kpc) for the SMC
following \citet{walborn00}. Our absolute magnitudes are slightly smaller than
Walborn, et al.'s (2000) due to lower extinction derived from \grid\  theoretical intrinsic
colors (typically, our color excesses are 0.02\,mag smaller).
The stellar luminosities were then obtained applying a bolometric correction from 
\citet{OS02}. This \grid\  relation, BC {\it vs.} \teff, is very similar to the
\citet{dekoter98} relation and to the empirical relation of \citet{chle91}.
The derivation of stellar radii is then straightforward.

\subsection{Microturbulent Velocity and Chemical Abundances}
\label{AbundSect}

We have used \grid\  model atmospheres with $Z/Z_\odot$= 1/5, and 1/10,
to determine the stellar metallicities, fitting the numerous iron lines in the
UV spectrum. Departures from scaled-solar
abundances have been considered for carbon and nitrogen. We have
recomputed model spectra with \synspec\   and CMFGEN assuming different C and N
abundances. Oxygen abundances cannot be derived reliably from our spectra:
\ion{O}{4}\,$\lambda$1338-43 strength is very sensitive to the
adopted microturbulent velocity, and \ion{O}{5}\,$\lambda$1371 is formed
in the stellar wind. Similarly to the \ion{O}{4} lines, \ion{S}{5}\,$\lambda$1502
is mostly useful to derive $\xi_{\rm t}$. We kept abundance ratios fixed
to the same scaled-solar value for all other species included in the model
atmospheres (O, Ne, Si, P, S).

\subsection{Wind parameters}
\label{WindParSect}

After the photospheric analysis, the
parameter space left to explore is greatly reduced. Using CMFGEN,
we derive in the next step the wind parameters (namely, \Mdot, \vinf\  and
$\beta$) keeping \teff\  within the error bars of the photospheric analysis.
Our analysis assumes an homogeneous wind and neglects clumping in the wind
at the present stage. We will however comment further on this point
in \S\,\ref{DiscSect}.

Initial estimates of the mass loss rates have been calculated with a theoretical
formula from \citet{vink01}. The mass loss rates are adjusted until we
match the observed spectra. The three stars with earlier spectral type exhibit 
P~Cygni line profiles in the UV spectrum. The primary diagnostic lines available
to determine the mass loss rate are the \ion{C}{4} resonance doublet,
the \ion{N}{5} resonance doublet, and \ion{O}{5}\,$\lambda$1371. The \ion{N}{5} 
lines are sensitive to the presence of shocks in the wind and a match to the
\ion{O}{5} line is often difficult to achieve. Other
potentially useful UV features are \ion{C}{3}\,$\lambda$1176,
\ion{O}{4}\,$\lambda$1338-43, \ion{He}{2}\,$\lambda$1640, and
\ion{N}{4}\,$\lambda$1718. In the optical, we have recorded spectra around
H$\alpha$ for these three stars. In all cases, H$\alpha$ is an
absorption line which does not significantly depart from a pure
photospheric profile. H$\alpha$ and other UV lines help nevertheless to set
limits on the mass loss rate. The spectrum of the three stars with a later
spectral type, NGC 346-113, NGC 346-487, and NGC 346-12, do not reveal any
P~Cygni line profiles, but only weak blue-shifted absorption in \ion{C}{4} and
\ion{N}{5} resonance lines. We could then only establish upper limits
on the mass-loss rate.

Simultaneously, we have to determine the exponent $\beta$ of the wind's velocity
law. The predicted line profiles are quite sensitive to the density in the
sonic region and thus depend on the adopted $\beta$. For weak stellar winds,
there is often a degeneracy in the determination of \Mdot\  and $\beta$ (see also
Hillier et al. 2003). We have adopted $\beta=1$ as the default value, but
we have still explored cases of slightly different $\beta$, $0.7\leq\beta\leq 1.5$.

Apart from \ion{N}{5}\,$\lambda$1240 in MPG~355, our spectra do not reveal any
saturated P~Cygni lines to measure the wind terminal velocity, \vinf.
Following \citet{prinja98}, we have thus used
narrow absorption components seen in MPG~324 and in MPG~368 spectra
to determine \vinf. We can only derive lower limits of the wind terminal
velocity for the three other stars. The maximum turbulent
velocity in the wind (Eq.~\ref{VturbEq}) is derived by fitting the blue edges of
the P~Cygni line profiles, while the minimum value has been set from the photospheric
analysis. Turbulence in the wind affects little the ionization structure of the models,
and is only necessary to achieve a match to the observed profiles.

We have adopted in a first step homogeneous wind models, but we have been unable
to match some of the line profiles (most notably, \ion{O}{5}\,$\lambda$1371, see
\S\ref{ResuSect}). We have therefore examined the influence of clumping in a second stage.

\subsection{Projected Rotational Velocities}
\label{VsiniSect}

\tlusty\  spectra are convolved with a rotational profile and a Gaussian
instrumental profile in order to compare them with observed spectra.
The projected rotational velocity \vsini\  is derived by matching relatively
weak iron lines, as well as \ion{C}{3}\,$\lambda$1176, \ion{O}{4}\,$\lambda$1338-43,
and \ion{S}{5}\,$\lambda$1502. The \ion{C}{3}\,$\lambda$1176 multiplet is
very useful because of its 6 close line components. In the STIS spectra, all
stars but NGC~346-355 (too hot to show strong \ion{C}{3} lines) show partial
or fully resolved multiplet components, indicating moderate ($V\sin i < 100$\,\kms)
to slow ($V\sin i\approx 20$\,\kms) rotational velocities. The projected rotational
velocities are determined with a typical precision of $\pm$10\,\kms.

Limited work exist so far on rotating stellar winds, because an exact calculation
of line profiles formed in an extended wind requires multi-dimensional model
atmospheres (see, e.g., Petrenz \& Puls 1996). Convolution with an
analytical profile is a reasonable approximation for photospheric lines only, as
long as the rotational velocity does not approach the break-up velocity.
In our case, most lines are formed in the quasi-static photosphere or in
the sonic region since the wind is relatively weak (very few lines even show
a P~Cygni profile). Moreover, rotational velocities remain small as indicated
by the \ion{C}{3}\,$\lambda$1176 multiplet. We believe therefore that
we may apply a rotational convolution to observer's frame spectra calculated
with CMFGEN with a relative safety.

\subsection{Interstellar Absorption Lines}
\label{ISMSect}

The UV spectra reveal two sets of narrow interstellar absorption lines, from
Galactic and SMC gas. A detailed modeling of these features is left to a future
paper. However, we have added Ly$\alpha$ interstellar contribution to the theoretical
spectra, in order to compare observed and predicted spectral
features near Ly$\alpha$ (e.g., \ion{Si}{3}\,$\lambda$1206, \ion{N}{5}\,$\lambda$1240).
We have varied the \ion{H}{1} column until a match to the observed Ly$\alpha$ profile
is achieved.
   
% ------------------------------------------------------------------------------

\section{Results of the Spectroscopic Analysis}
\label{ResuSect}

We discuss in this section the properties, the results and uncertainties of the
analysis of each individual star. A general discussion about the sample as a whole,
of the astrophysical significance of the results, and a comparison to theoretical
predictions is left to the next section.
% Figures~1 and 2 show the best \tlusty\ 
% and CMFGEN models for the full observed spectral range in the UV and in the
% optical and for all six stars simultaneously, so that trends are readily
% apparent as well as potential issues with the adopted atomic data.\footnote{Fig.
% 1 and 2 are only published electronically due to space consideration, but some
% line profile fits are presented in the next figures.}
\subsection{NGC 346 MPG 355}
\label{MPG355Sect}

This star has long been classified as an O3\,III~((f*)) star \citep{massey89, walborn00},
based on the absence of \ion{He}{1}\,$\lambda$4471
in its spectrum. Recently, \citet{walborn02} have reexamined the earliest O-type stars
and have introduced two new spectral classes, O2 and O3.5, to distinguish
stars having no or very weak \ion{He}{1}\,$\lambda$4471, but different \ion{N}{3-v}
lines. MPG~355 exhibits \ion{N}{5}\,$\lambda$4603-19 in absorption, strong
\ion{N}{4}\,$\lambda$4058 in emission, and no \ion{N}{3}\,$\lambda$4640. 
This star belongs thus to the newly defined O2 class, and is classified O2\,III~(f*).
\citet{taresch97} used similarly the nitrogen ionization balance
to derive the effective temperature of the O3If star, HD~93129A.

\citet{haser98} have analyzed a low resolution (7\,\AA) HST/FOS spectrum
of MPG~355. We compare their findings to our new results based on higher resolution
spectra recorded with HST/STIS.

The spectral type, the helium and nitrogen lines indicate that this star is unusually hot.
We found that a \tlusty\  model atmosphere with \teff = 52\,500\,K, $\log g$=4.0, and
a metallicity 1/5 the solar value, provides the best match to the observations.
This model atmosphere reproduces also well the \ion{Fe}{4-vi} line strength ratios in
the UV. The \ion{C}{3}\,$\lambda$1176 lines are very weak and can only be used to set
a lower limit to \teff. A slightly lower effective temperature, \teff\ $\approx$ 50\,000\,K,
is favored by \ion{S}{5}\,$\lambda$1502, while the fit to \ion{He}{2}\,$\lambda$1640, 4686
is improved at higher \teff\  (54-55,000\,K). We estimate the uncertainties to be $\pm$2500\,K
and $\pm$0.1\,dex on \teff\  and $\log g$, respectively. Within the quoted uncertainties, our
results agree with \citet{haser98} values.

Simultaneously, we have derived a microturbulent velocity, $\xi_{\rm t}$=25\,\kms, from a fit
to \ion{S}{5}\,$\lambda$1502, while assuming a sulfur abundance, S/S$_\odot$=0.2.
The \ion{O}{4}\,$\lambda$1338-43 and the Fe lines are also well matched with this microturbulent
velocity, assuming 1/5 times the solar abundances. The Fe lines can alternatively be matched with
a lower abundance, one tenth of the solar value, while increasing the turbulent velocity to
a supersonic value ($\xi_{\rm t}\approx$~35\,\kms); however, the fit to other lines, e.g.
\ion{C}{4}\,$\lambda$1169 or \ion{S}{5}\,$\lambda$1502, is lost. A microturbulent velocity as
large as 25\,\kms\  is of similar order than the photospheric sound speed in MPG~355. Although
a quite large value, it is unlikely that this is a spurious result. Indeed, such large microturbulence
is required to match lines of different species and is therefore not a result of shortcomings
of the adopted model atoms. Furthermore, CMFGEN models also require a microturbulent velocity
of the same order to match the observed spectrum. This microturbulence mimics non-thermal motions
in the atmosphere. Results from CMFGEN models show that this large microturbulence
is not a spurious artefact
of static model atmospheres which neglect the desaturation of spectral lines 
by a macroscopic velocity field -- the stellar wind, as was suggested by \citet{kudr92}.
Denser winds are necessary for the latter to become effective (i.e. lines need to be formed
in the sonic region, but they will then show a blue asymmetry).

A match to the observations is achieved
after including a rotational convolution, with $V\sin i$=110\,\kms.

In the ultraviolet, the carbon abundance can only be determined by fitting \ion{C}{4}\,$\lambda$1169.
Only an upper limit can be set with \ion{C}{3}\,$\lambda$1176, while the \ion{C}{4}\,$\lambda$1549
is formed in the wind and its strength depends on the adopted mass loss rate and other
wind parameters. A carbon abundance, C/C$_\odot$=0.2, provides a good match
to \ion{C}{4}\,$\lambda$1169. The $3s-3p$, \ion{C}{4}\,$\lambda$5801-12 lines are predicted in
emission by this hot model atmosphere and are observed in emission. This doublet is well fitted
with the adopted carbon abundance. Moreover, \ion{C}{4}\,$\lambda$5801-12 emission brings further
support to our adopted effective temperature. \tlusty\  \grid\  models indeed predict
that this doublet turns into emission at \teff\,$\geq$\,50\,000\,K, with stronger emission
at higher effective temperatures. At \teff\,$\leq$\,47\,500\,K, these lines are predicted in
absorption.

From \ion{N}{4}\,$\lambda$4058 and \ion{N}{5}\,$\lambda$4603-19, we have derived a solar
nitrogen abundance, showing thus a significant nitrogen enrichment at the stellar surface.
Even with this overabundance, the model predicts no visible \ion{N}{3}\,$\lambda$4640,
as observed. This nitrogen abundance also provides a good fit to \ion{N}{4}\,$\lambda$1718,
while the \ion{N}{3}\,$\lambda$1182-84 lines remain very weak. The \ion{N}{5} resonance doublet
cannot be used to determine a reliable abundance since the \ion{N}{5} density in the wind
might be strongly affected by shocks and the related X-ray emission.

Oxygen abundance cannot be derived reliably: \ion{O}{4}\,$\lambda$1338-43 lines are most
sensitive to the adopted microturbulent velocity, while \ion{O}{5}\,$\lambda$1371 is formed
in the wind (see further discussion below). We have thus assumed by default a scaled-solar
value comparable to the derived metallicity (1/5 solar).

\citet{haser98} have similarly found a stellar metallicity of 1/5 the solar
metallicity, based on spectrum synthesis analysis of the low resolution HST/FOS spectrum
and indirect hydrodynamical arguments. We stress however that our determination
is more robust because the higher resolution of STIS allows us to disentangle the numerous line
blends in the UV spectrum. We estimate that the accuracy of our abundance determinations is typically
$\pm$0.2\,dex. Already from line strength consideration (\ion{N}{5}\,$\lambda$1240 {\it vs.}
\ion{C}{4}\,$\lambda$1549), Walborn et~al. (1995, 2000) recognized
that MPG~355 has a surface composition very likely enriched in nitrogen, which was confirmed
by Haser et~al. analysis. Our results indicate an even larger nitrogen enrichment.
On the other hand, Haser et~al. found a very low carbon abundance, C/C$_\odot$=0.02,
claiming even that this value had to be an upper limit. This result was based on
the \ion{C}{4}\,$\lambda$1549 resonance doublet, for which Haser et~al. estimated a
contribution from shocks in the wind. They were unable to find any consistent solution for
several wind lines without adopting this very low carbon abundance. The \ion{C}{4}\,$\lambda$1169
line in our STIS spectrum cannot be reconciled at all with Haser et~al. low carbon abundance.

\begin{figure}
\figurenum{1}
\plotone{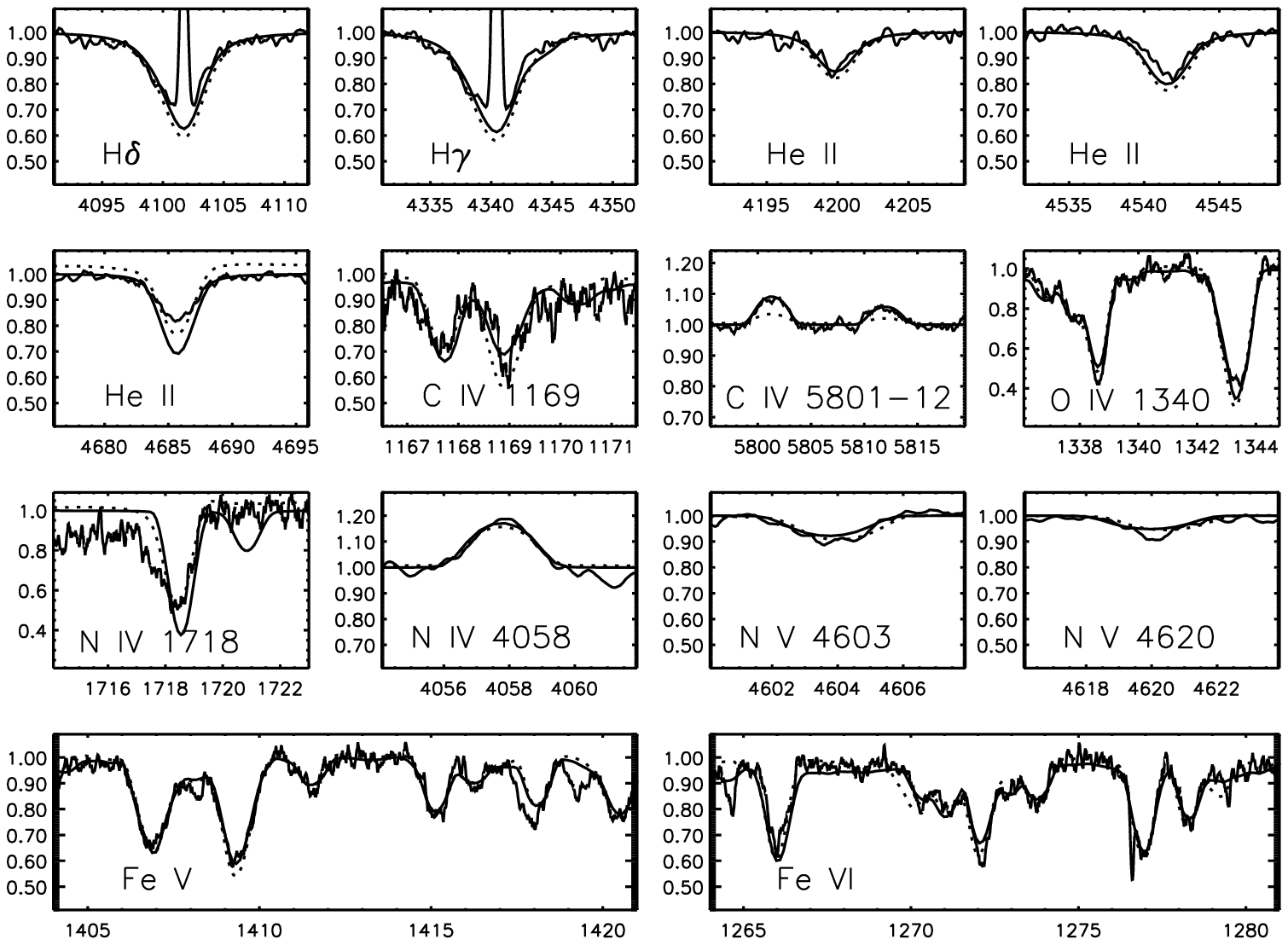}
\caption{Best fits to MPG 355 ``photospheric'' lines used to derive the stellar parameters,
  \teff=52500\,K, $\log g=4.0$, and surface chemical composition, $Z/Z_\odot=0.2$
   (full line: \tlusty, dotted line: CMFGEN). \label{P355Fig}}
\end{figure}

Fig.~\ref{P355Fig} displays the best \tlusty\  and CMFGEN model fits to these lines formed
in the ``quasi-static'' photosphere.

Using the parameters and abundances derived with \tlusty, we have then constructed a number
of wind models with CMFGEN. We achieve a good agreement between \tlusty\  and CMFGEN models
for lines of weak and moderate strength formed in the pseudo-photosphere.
Wind parameters are derived from the strongest lines. The terminal velocity and maximum
turbulent velocity in the wind are set from the only saturated lines, \ion{N}{5}\,$\lambda$1240.
We have adopted a terminal velocity, \vinf = 2800\,\kms, and a maximum turbulent velocity
in the wind of 160\,\kms.
The mass loss rate and wind acceleration parameter, $\beta$, are derived by fitting
\ion{C}{4}\,$\lambda$1549 and H$\alpha$. We found a mass loss rate,
\Mdot = 2.5\,10$^{-6}$\,\msolyr, for a wind without clumps, and $\beta=0.8$, in order
to reproduce \ion{C}{4} emission without filling the H$\alpha$ core too much.
We can exclude slow acceleration winds, $\beta\geq 1.5$, for which we cannot achieve
a consistent fit of \ion{C}{4}\,$\lambda$1549 and H$\alpha$, given the adopted carbon
abundance. Emission due to incoherent electron scattering in the wings of
\ion{He}{2}\,$\lambda$1640, 4686\footnote{The extent of \ion{He}{2}\,\lb4686 wings remains
ill-defined due to difficulties in defining an ``exact'' continuum in the echelle spectrum.}
is predicted somewhat too strong.
A higher effective temperature (55\,000\,K) would solve this problem.
   
\ion{N}{5}\,$\lambda$1240 is also predicted somewhat
too strong. Furthermore, our wind model does not reproduce well the
observed asymmetry of \ion{N}{4}\,$\lambda$1718 (extended blue absorption), contrary to the
case of the cooler stars, MPG~324 and MPG~368. Our adopted wind model predicts thus 
too high a nitrogen ionization in the wind of MPG~355. On the other hand, the photospheric
\ion{N}{4}/\ion{N}{5} ionization balance reproduces well the lines in the blue spectrum.
The  theoretical soft X-ray flux ($\lambda<160$\,\AA, \ion{N}{4} ionization edge) is
of correct order in the photosphere and in the sonic region but is still too strong in the wind,
even though we have neglected the possible influence of shocks. We note that
\citet{haser98} similarly predicted \ion{N}{5}\,$\lambda$1240 too strong.

\begin{figure}
\figurenum{2}
\plotone{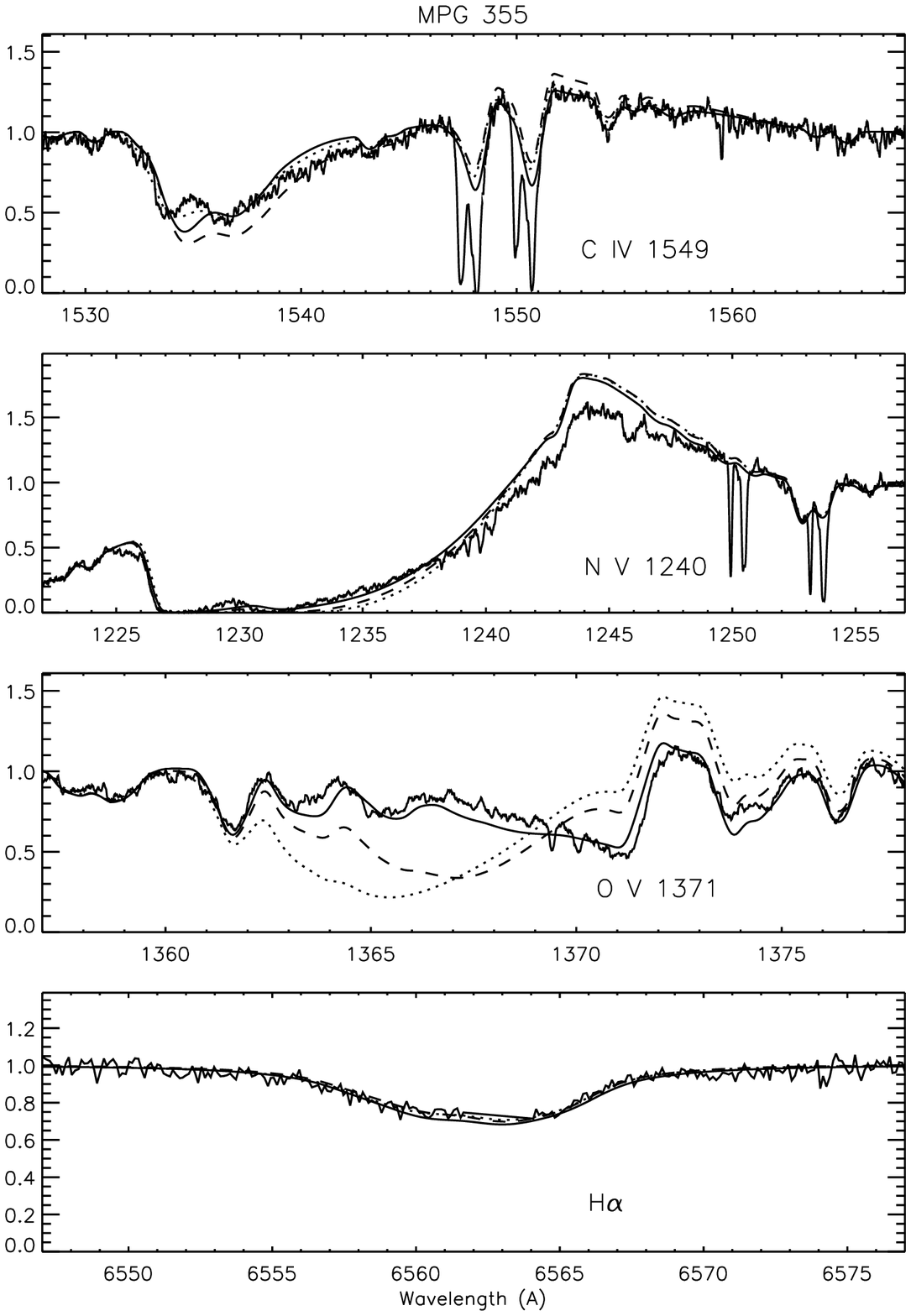}
\caption{Influence of clumping on MPG 355 wind lines. A comparison of three models with
  different clump volume filling factors: no clumps, $f_\infty=1.0$ (dotted);
  $f_\infty=0.1$ (dashed);
  and best fit, $f_\infty=0.01$ (full line). The mass loss rates have been adjusted to match
  the \ion{C}{4}\,$\lambda$\,1549 lines, and are \Mdot = 2.5\,\evi, 7.9\,\evii, and
  1.8\,\evii\,\msolyr, respectively. The blue side of \ion{N}{5}\,$\lambda$1240 is affected
  by interstellar Ly\,$\alpha$ absorption. For display purposes, the core of H$\alpha$ was 
  cut off to remove nebular contributions. \label{W355Fig}}
\end{figure}

\ion{O}{5}\,$\lambda$1371 reveals a problem encountered in all previous analyses
of O stars that show \ion{O}{5}\,$\lambda$1371:
the model wind atmospheres predict too strong absorption at moderate velocities,
too little absorption at low velocities, and too strong emission. We have thus examined the
influence of clumping on this line. We have calculated several models with different clump
volume filling factors, adjusting the mass loss rate so that we keep a good fit to
\ion{C}{4}\,$\lambda$1549. H$\alpha$ core is somewhat filled by wind emission and  provides
a further constraint on the mass loss rate. Fig.~\ref{W355Fig} shows that we can achieve a
good match of \ion{O}{5}\,$\lambda$1371 when assuming a very small filling factor, $f_\infty=0.01$,
with a mass loss rate, \Mdot = 1.8\,10$^{-7}$\,\msolyr. A larger filling factor, $f_\infty=0.1$,
and a lower oxygen abundance (by a factor 4) does not fit the observed profile as well as the
very small $f_\infty$ though it already improves over the case of homogeneous winds.
Although the larger clumping factor seems more reasonable and the \ion{O}{4}\,$\lambda$1338-41
lines are still matched with the lower abundance (these lines are saturated and are thus quite
insensitive to changes in abundance), we do not find any compelling argument to adopt an oxygen
abundance a factor 3 lower than the nebular abundance while carbon is not depleted as well
(as carbon depeletion is expected first for CNO-cycle processed material).
The sharp transition between the absorption and emission components indicates that clumps
start forming at low velocities, $v\approx 30$\,\kms, just above the sonic point. The other
wind line profiles are little changed relative to profiles predicted by an homogeneous wind model.

We have explored a number of other possibilities in order to reproduce \ion{O}{5}\,$\lambda$1371.
The key point consists in predicting the correct \ion{O}{4}/\ion{O}{5} ionization structure
in the vicinity of the sonic point. To alter ionization, we have considered X-ray emission
from shocks in the wind ($L_X/L_{\rm bol}=2\,10^{-7}$),
adiabatic cooling, larger model atoms, and alternate recombination
rates \citep{nahar99}. All these changes resulted in essentially no difference in the synthetic
spectrum.

Alternatively, \citet{dekoter98} argued that O3 stars are
cooler (40\,000\,K $\leq$ \teff\  $\leq$ 46\,000\,K), based on the \ion{O}{4}/\ion{O}{5}
line strength ratio. These low temperatures are not supported by our other \teff\  estimators.
Moreover, the mismatch of the \ion{O}{5}\,$\lambda$1371 line is still qualitatively the same,
revealing a problem with the wind models at low and moderate velocities where the validity of
the Sobolev approximation (assumed by de Koter's ISA-WIND program) is not established. We
stress however that the critical issue is clumping, and not the Sobolev approximation,
as demonstrated by the failure of homogeneous wind models (see Fig.~\ref{W355Fig},
and also Fig.~9 in Haser et~al.) to reproduce the observed line profile.

A full examination of the far-UV spectrum and of the visible spectrum reveals that only few other
lines are influenced by the stellar wind, even for a star with such an early spectral type and
a moderate wind (\Mdot\ $\approx$ few $10^{-6}$\,\msolyr). These lines include
\ion{O}{4}\,$\lambda$1338-43, \ion{N}{4}\,$\lambda$1718, and few strong Fe lines, whose cores
are filled a little by wind emission.   
Otherwise, we have a very good consistency between \tlusty\  and CMFGEN model spectra.
This agreement shows that we can use NLTE photospheric models to derive stellar
parameters and abundances, even for the hottest O dwarfs with moderate winds.
% as long as we accept to adopt a relatively large microturbulent velocity.

\subsection{NGC 346 MPG 324}
\label{MPG324Sect}

We are able to match the Balmer line series and the \ion{He}{2}
lines with a \tlusty\  model atmosphere, assuming \teff\  = 40\,000\,K
and $\log g=4.0$. On the other hand, the \ion{He}{1} lines are virtually
absent which suggests an effective temperature of about 45\,000\,K. 
However, the visible spectrum is contaminated with nebular
emission, suggesting that the weakness of the \ion{He}{1} lines
is mostly due to line filling by this emission.
\citet{walborn02} showed that the different ionization stages
of nitrogen can be used as a temperature indicator, but the nitrogen
lines are hardly seen in the visible spectrum of MPG~324 and therefore
cannot be used for this purpose. Based on the visible
spectrum only, our estimates support earlier results from
\citet{puls96}, who found \teff\ = 40\,000\,K and
$\log g = 3.7$.

\begin{figure}
\figurenum{3}
\plotone{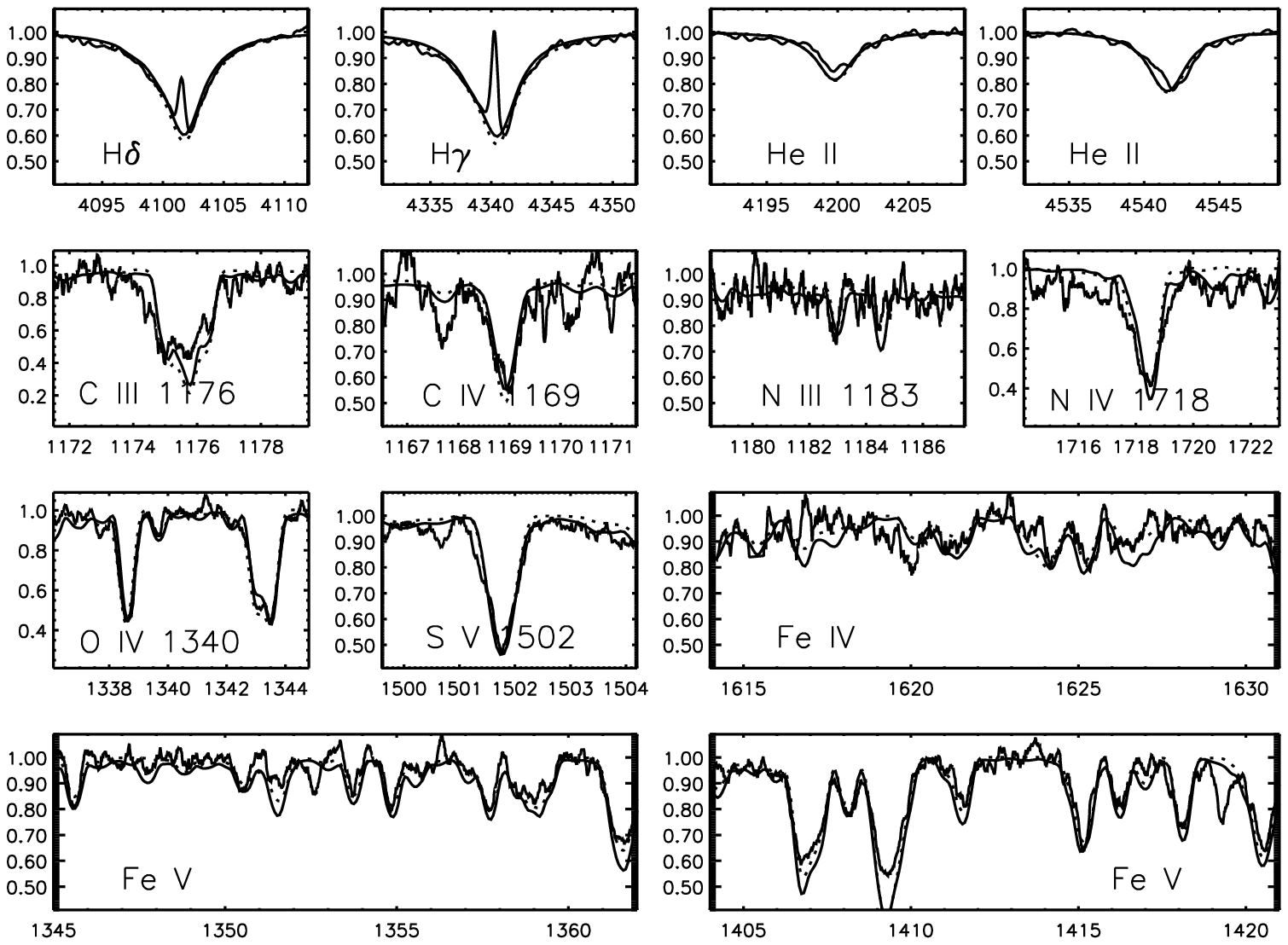}
\caption{Best fits to MPG 324 photospheric lines used to derive the stellar parameters,
  \teff=41500\,K, $\log g=4.0$, and surface chemical composition, $Z/Z_\odot=0.2$
   (full line: \tlusty, dotted line: CMFGEN). \label{P324Fig}}
\end{figure}

The UV spectrum provides additional temperature criteria and 
\tlusty\  and CMFGEN model atmospheres lead to exclude the higher
temperatures suggested by the weakness of the \ion{He}{1} lines.
Using the \ion{C}{3} to \ion{C}{4} and the
\ion{Fe}{4}/\ion{Fe}{5}/\ion{Fe}{6} ionization balances,
our analysis yields a best overall match when adopting
\teff\  = 41\,500\,K and $\log g=4.0$. Uncertainties are
estimated at $\pm$2\,000\,K in temperature and $\pm$0.1\,dex
in gravity. We achieve consistent results between the UV
and visible spectra, as well as between \tlusty\  and CMFGEN
models. Fig.~\ref{P324Fig}  displays the best fits to photospheric
lines.

We have derived a moderate apparent rotational velocity,
$V \sin i$ = 70\,\kms, from the partly resolved line structure
of the \ion{C}{3}\,$\lambda$1176 multiplet. This is slightly
lower than a previous estimate listed by \citet{walborn00}.
We then fix the microturbulence velocity, $\xi_{\rm t}=15$\,\kms,
by matching \ion{O}{4}, \ion{S}{5}, and Fe lines, assuming
a scaled-solar model with $Z/Z_\odot=0.2$. Alternatively, we
can adopt a higher microturbulence velocity, $\xi_{\rm t}=20$\,\kms,
and a lower metallicity, $Z/Z_\odot=0.1$. The latter fit is not as good,
but provides an estimate of the typical uncertainty on
abundances. Although we cannot derive oxygen and sulfur abundances
independently to the microturbulence value, we have examined
the case of a lower oxygen abundance because the adopted abundance,
O/O$_\odot$=0.2, provides a poor fit to \ion{O}{5} \lb1371. We cannot
reduce the oxygen abundance by much without increasing the microturbulent
velocity to unrealistic values. Moreover, we remain unable to reproduce
the \ion{O}{5} \lb1371 line with homogeneous wind models.

The carbon abundance was derived using \tlusty\  and \synspec\
models to fit simultaneously \ion{C}{3} \lb1175 and \ion{C}{4}
\lb1169, using the parameters previously derived. We have also
constrained the carbon abundance from the absence of \ion{C}{3}
\lb4647-51 in the visible spectrum (neither emission nor absorption).
We found that the SMC nebular abundance, C/C$_\odot$=0.06
\citep{venn99}, results in a good fit.

The nitrogen abundance,
N/N$_\odot$=0.2, is derived from the \ion{N}{3} \lb1182-84 lines.
The other nitrogen lines are not as sensitive as these lines to
derive the nitrogen abundance. The \ion{N}{5} resonance doublet
is quite sensitive to the adopted wind parameters, the \ion{N}{4}
\lb1718 is only marginally sensitive to the abundance, while the lines
in the visible spectrum are very weak given the signal-to-noise ratio
of the data. While it might appear that the derived abundance
remains subject to some uncertainty, we can exclude low
abundances, N/N$_\odot < 0.1$, for which we cannot match the
UV \ion{N}{4} and \ion{N}{5} lines, as well as large abundances,
N/N$_\odot > 0.5$ which would yield nitrogen emission lines in the
visible.

\begin{figure}
\figurenum{4}
\plotone{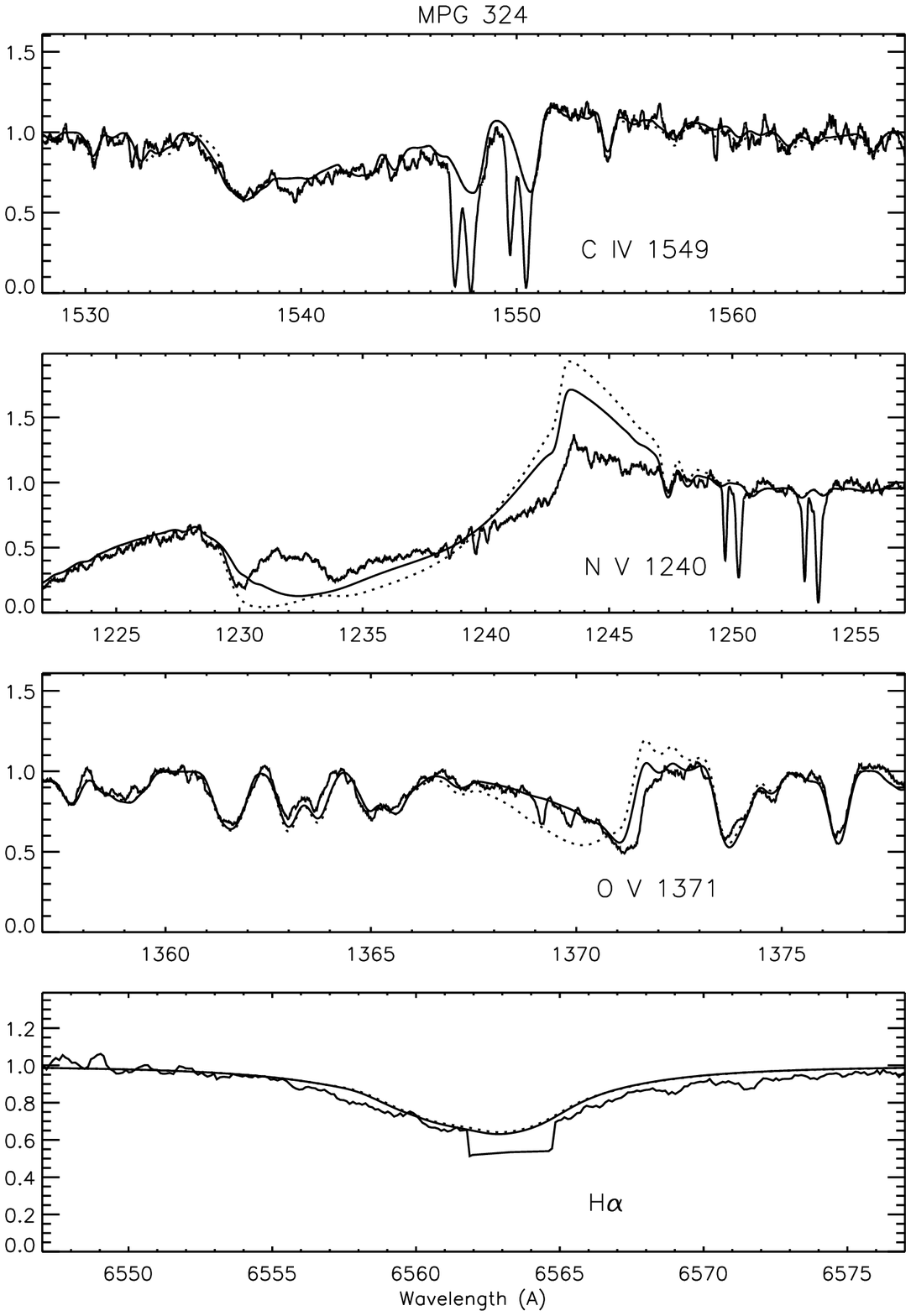}
\caption{Influence of clumping on MPG 324 wind lines. A comparison of two models with
  different clump volume filling factors: no clumps, $f_\infty=1.0$ (dotted);
  and best fit, $f_\infty=0.1$ (full line). The mass loss rates have been adjusted to match
  the \ion{C}{4}\,$\lambda$\,1549 lines, and are \Mdot = 2.7\,\evii, and
  1.0\,\evii\,\msolyr, respectively. The blue side of \ion{N}{5}\,$\lambda$1240 is affected
  by interstellar Ly\,$\alpha$ absorption. For display purposes, the core of H$\alpha$ was 
  cut off to remove nebular contributions. \label{W324Fig}}
\end{figure}

From \ion{N}{5}\,\lb1240, we have measured a wind terminal
velocity, \vinf\ = 2300\,\kms. This is in good agreement with
a value derived from narrow absorption components. Therefore,
even though the wind lines are not totally saturated, this value
should likely not be much lower than the actual terminal velocity.
The mass loss rate and the $\beta$ exponent of the wind velocity
law have been determined fitting the few available P~Cygni
profiles, in conjunction with other wind sensitive lines, like
\ion{He}{2}\,\lb1640\ or \ion{N}{4}\,\lb1718 in the UV and \halpha\ 
in the visible. For an homogeneous wind, we find \Mdot\ =
2.7$\times 10^{-7}$ \msolyr, and $\beta$ = 1.0. Our estimate of the mass
loss rate is consistent with an upper limit derived by
\citet{puls96}. The fits to the wind lines of MPG~324 are
displayed in Fig.~\ref{W324Fig}. Similarly to MPG~355, the homogeneous
wind models fail to reproduce the \ion{O}{5}\,\lb1371 line profile.
A much improved fit is achieved with a clumped wind model assuming a
volume filling factor, $f_\infty=0.1$, and a mass loss rate, \Mdot\ =
1.0$\times 10^{-7}$ \msolyr.

\subsection{NGC 346 MPG 368}
\label{MPG368Sect}

\begin{figure}
\figurenum{5}
\plotone{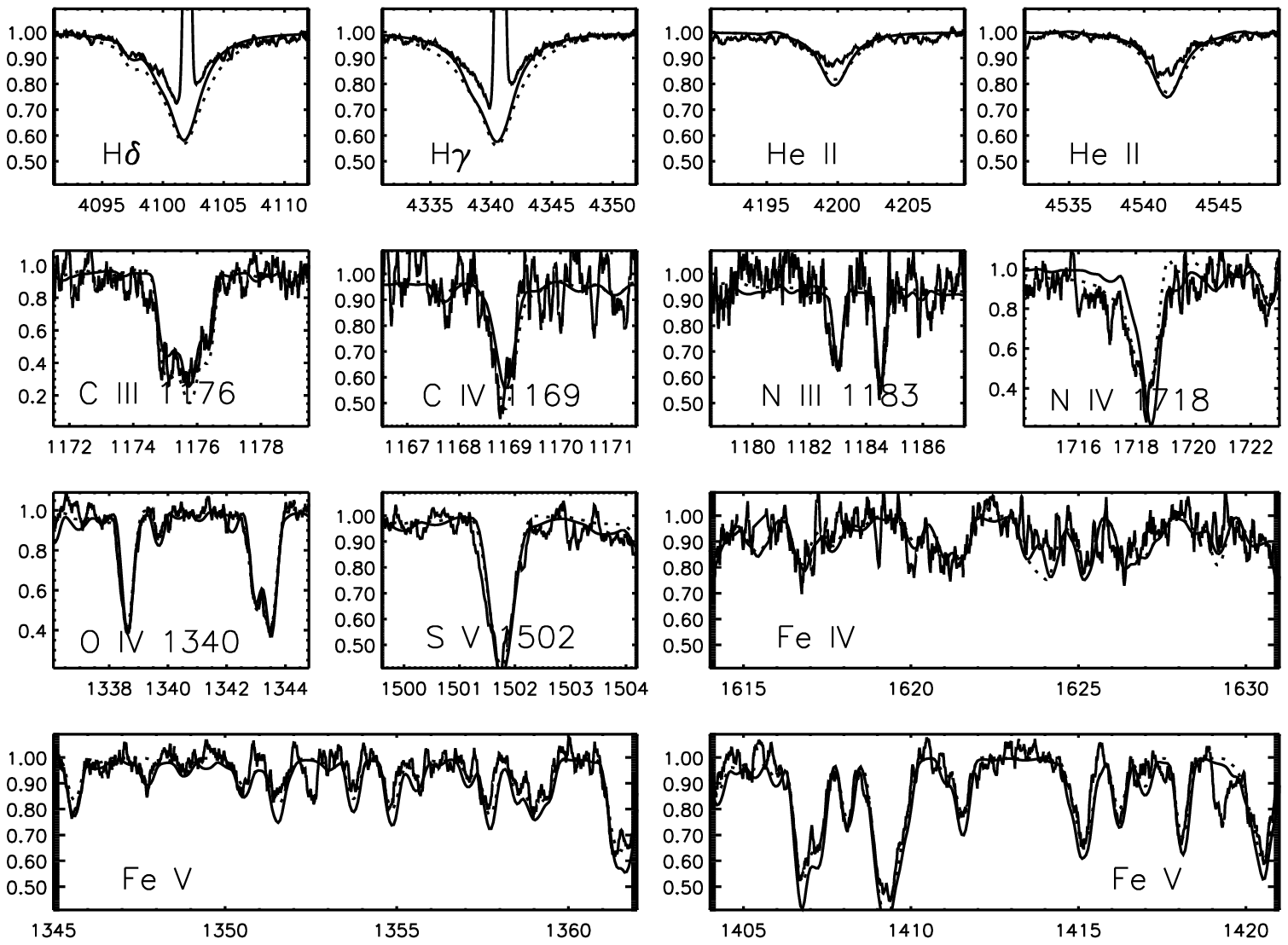}
\caption{Best fits to MPG 368 photospheric lines used to derive the stellar parameters,
  \teff=40000\,K, $\log g=3.75$, and surface chemical composition, $Z/Z_\odot=0.2$
   (full line: \tlusty, dotted line: CMFGEN). \label{P368Fig}}
\end{figure}

\begin{figure}
\figurenum{6}
\plotone{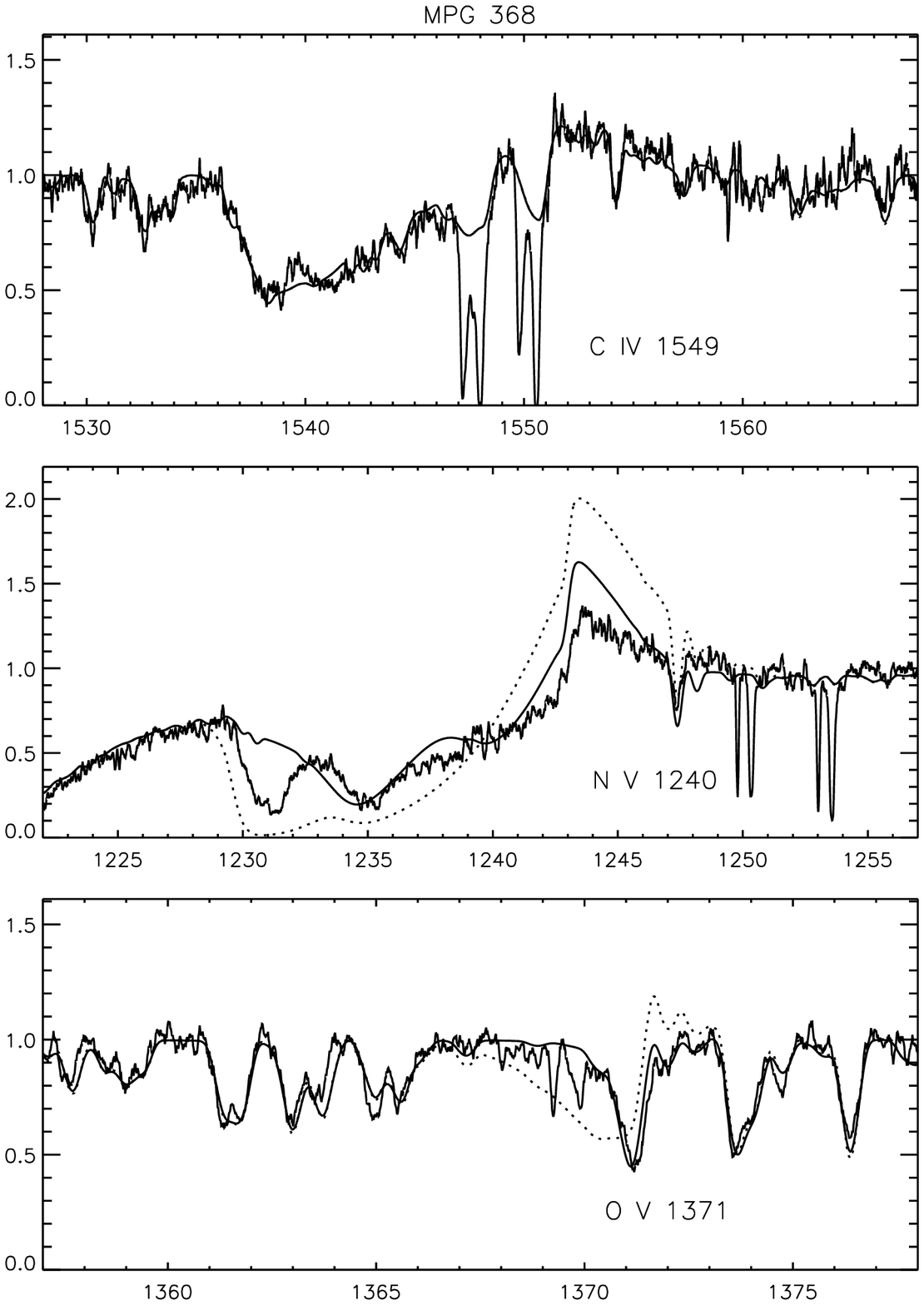}
\caption{Influence of clumping on MPG 368 wind lines. A comparison of two models with
  different clump volume filling factors: no clumps, $f_\infty=1.0$ (dotted);
  and best fit, $f_\infty=0.05$ (full line). The mass loss rates have been adjusted to match
  the \ion{C}{4}\,$\lambda$\,1549 lines, and are \Mdot = 1.5\,\evii, and
  5.7\,\eviii\,\msolyr, respectively. The blue side of \ion{N}{5}\,$\lambda$1240 is affected
  by interstellar Ly\,$\alpha$ absorption. \label{W368Fig}}
\end{figure}

\begin{figure}
\figurenum{7}
\plotone{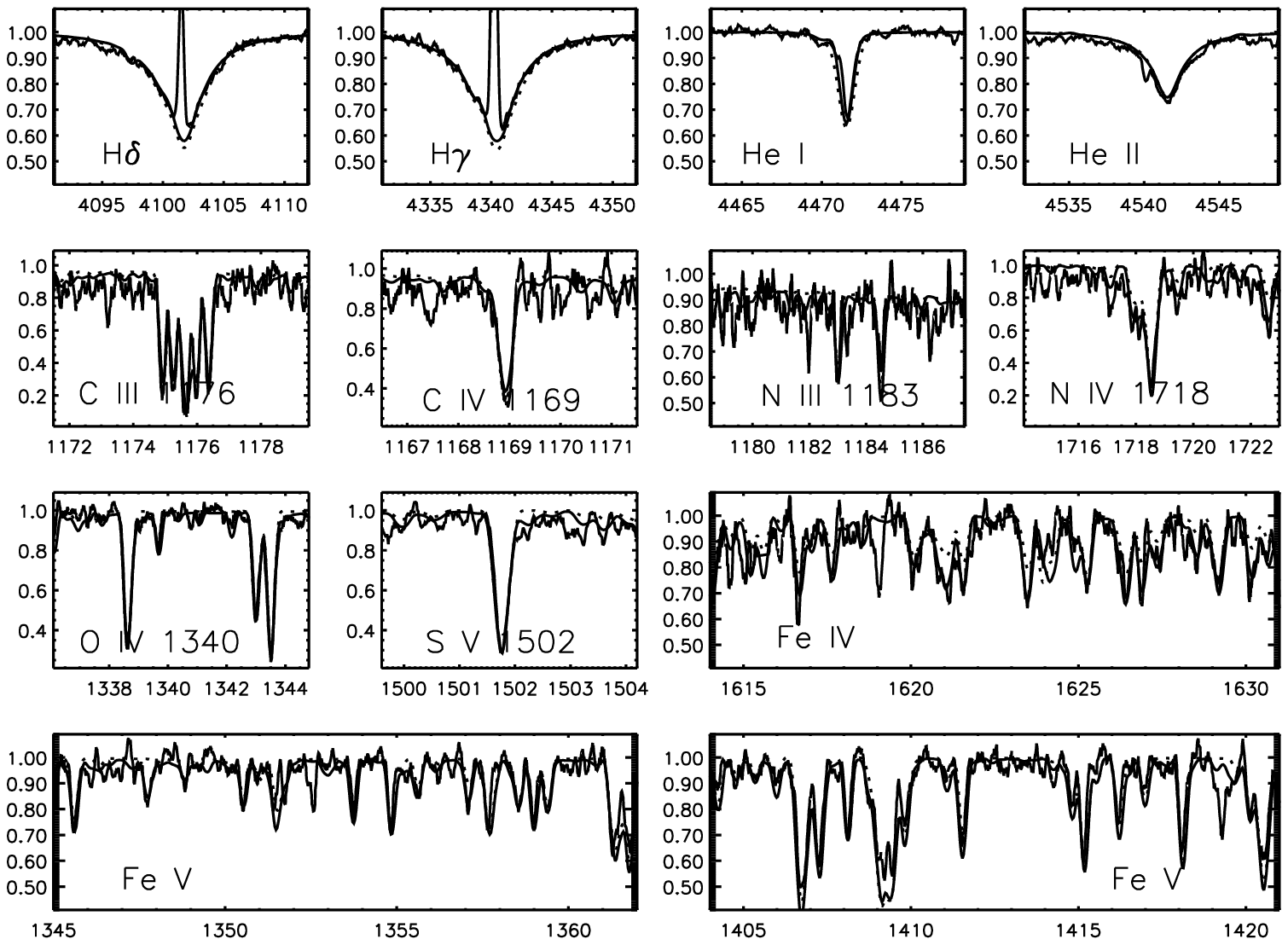}
\caption{Best fits to MPG 113 photospheric lines used to derive the stellar parameters,
  \teff=40000\,K, $\log g=4.0$, and surface chemical composition, $Z/Z_\odot=0.2$
   (full line: \tlusty, dotted line: CMFGEN). \label{P113Fig}}
\end{figure}

As indicated by the spectral type, strong spectral similarities
exist between MPG~368 and MPG~324. We have thus started our analysis
of MPG~368 assuming MPG~324 stellar parameters and only minor changes
were required to fit MPG~368 spectrum. We discuss our results in a
comparative way to MPG~324; the same spectral regions are thus displayed
in Fig.~\ref{P368Fig} and Fig.~\ref{W368Fig}.

The analysis of the photospheric spectrum in the optical is hampered
by strong H and He nebular emission, resulting in some uncertainty 
in deriving the effective temperature. However, like MPG~324, the carbon
and iron ionization balances from UV lines allow us to derive
reliably an effective temperature, \teff\  = 40\,000\,K. Balmer
line wings are slightly narrower than in MPG~324, resulting in
a best fit for $\log g=3.75$. The \ion{He}{2} lines are not matched
as well as in MPG~324 spectrum; the adopted model
atmosphere yields too deep \ion{He}{2} lines compared to the
observed line profiles. \ion{O}{4}, \ion{S}{5}, and Fe lines,
are matched adopting the same microturbulent velocity,
$\xi_{\rm t}=15$\,\kms, the same metallicity, $Z/Z_\odot=0.2$, and
a rotational velocity, $V \sin i=60$\,\kms.

The carbon abundance is derived in matching \ion{C}{3}\,\lb1176
and \ion{C}{4}\,\lb1169. The models should also not predict
strong \ion{C}{3}\,\lb4647-51 emission that is not observed.
We find a carbon abundance, C/C$_\odot$ = 0.06, that is consistent
with the SMC nebular abundance, and similar to MPG~324 abundance.
The nitrogen
lines are clearly stronger in MPG~368 (e.g., \ion{N}{3}\,\lb1182-84,
hint of an emission in \ion{N}{3}\,\lb4634-40), thus indicating
that the surface of MPG~368 has been enriched in nitrogen.
We have adopted, N/N$_\odot$ = 0.6. The nitrogen enrichment
of MPG~368 compared to MPG~324 is consistent with its being more
evolved (indicated by a lower $\log g$) than MPG~324.

The wind lines of the two stars are also very similar, and we
have derived with CMFGEN similar parameters. Consistent with
the slightly lower effective temperature and luminosity, we
have obtained a slightly lower mass loss rate, \Mdot\ =
1.5$\times 10^{-7}$ \msolyr, lower terminal velocity,
\vinf\ = 2100\,\kms, and $\beta$ = 1.0. Abundances are the
same than those adopted from the photospheric analysis. The nitrogen
enrichment results in a stronger asymmetry of \ion{N}{4}\,\lb1718
compared to the case of MPG~324, and our final model reproduces
well the extended blue absorption.

Like MPG~355 and MPG~324, we are able to fit \ion{O}{5}\,\lb1371
when allowing for clumps, while such a match cannot be achieved with
homogeneous wind models. We achieve a best fit with a slightly
lower volume filling factor, $f_\infty=0.05$, and a mass loss rate,
\Mdot = 5.7\,\eviii\,\msolyr. The fit to \ion{N}{5}\,\lb1240 is also
clearly better with the clumped wind model, though the emission is still
predicted too strong without having yet accounted for the overionization
due to shocks. 
The detailed profile of the \ion{N}{5} lines may indicate that
clumping follows a distribution that is more complicated than
assumed, as suggested by the strong absorption component seen
in MPG~324 and MPG~368 near terminal velocity.

\subsection{NGC 346 MPG 113}
\label{MPG113Sect}

This star has been classified as OC6\,Vz by
\citet{walborn00}. They noted the presence of weak
\ion{C}{3}\,$\lambda$\,4647-50 emission lines without
\ion{N}{3}\,$\lambda$\,4634-40, indicating a low N/C
abundance ratio. Furthermore, the large relative strength
of \ion{He}{2}\,$\lambda$\,4686, the weak \ion{He}{1} lines,
and the extreme weakness of the UV wind lines, lead them to
assign a Vz class to MPG~113 which suggests thus a
very young age.

Our analysis yields an excellent match to the hydrogen and helium
spectrum with the \tlusty\ model, (\teff, $\log g$) = (40\,000\,K, 4.0).
This model agrees also very well with the UV spectrum once we have
fixed the turbulent microturbulent velocity, $\xi_{\rm t}$\,=\,10\,kms,
with \ion{O}{4} and \ion{S}{5} lines. The derived metallicity
from the numerous UV lines is $Z/Z_\odot=0.2$. All fine-structure
components of \ion{C}{3}\,$\lambda$\,1176 are resolved, indicating
a low projected rotational velocity.

The carbon abundance is derived from \ion{C}{3}\,$\lambda$\,1176 and
\ion{C}{4}\,$\lambda$\,1169. The \ion{C}{3}\,$\lambda$\,4647-51
recombination lines are also very sensitive to the adopted carbon 
abundance. The UV lines are well matched, adopting C/C$_\odot = 0.2$,
though they are not very sensitive. A lower abundance, C/C$_\odot = 0.1$,
does not deteriorate the fit much.
The \tlusty\  \grid\  model does not reproduce \ion{C}{3}\,$\lambda$\,4650
in emission, due to missing high-excitation levels in the \ion{C}{3} model
atom that results in underestimating C$^{3+}$ to C$^{2+}$ recombination rate.
We calculated a new model with a larger \ion{C}{3} model atom, predicting
these lines slightly in emission. This issue needs to be reexamined, but
we cannot use these lines to derive an abundance with current \tlusty\  models. 
The CMFGEN model includes a better model atom and satisfactorily matches 
the observed emission with the lower carbon abundance, C/C$_\odot = 0.1$.
The higher value yields \ion{C}{3} lines that are clearly too strong.
We have therefore adopted a carbon abundance,  C/C$_\odot = 0.1$.

The nitrogen abundance is set by \ion{N}{3}\,$\lambda$\,1182-84 and
\ion{N}{4}\,$\lambda$\,1718. A good fit is obtained with an abundance,
N/N$_\odot = 0.2$. With this abundance, only very weak \ion{N}{3}
lines (4097, 4640\,\AA) are predicted, as observed. The UV lines
provide therefore a N/C abundance ratio slightly above solar, contrary to
\citet{walborn00} suspicion of a low N/C ratio. Actually,
nitrogen is not as much enriched as in other stars of our sample,
which might have lead them to classify this star as an OC6 star. Conversely,
this might suggest that many ``normal'' O6\,V stars might be in fact
nitrogen-rich.

\begin{figure}
\figurenum{8}
\plotone{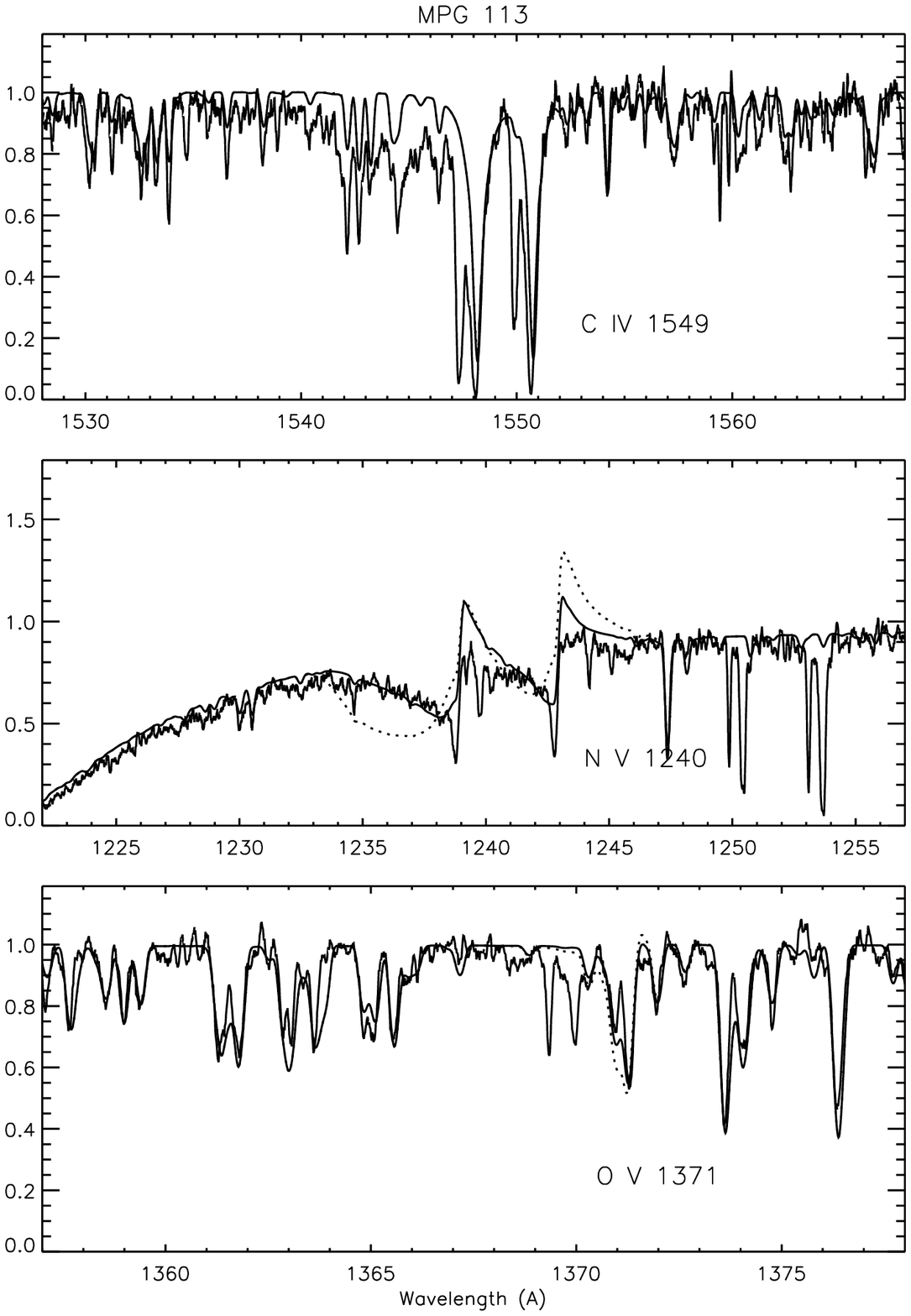}
\caption{Influence of clumping on MPG 113 wind lines. A comparison of two models with
  different clump volume filling factors: no clumps, $f_\infty=1.0$ (dotted);
  and best fit, $f_\infty=0.1$ (full line). The mass loss rates
  are \Mdot = 3\,\eix, and 1\,\eix\,\msolyr, respectively. The \ion{C}{4} lines 
  predicted by the homogeneous and clumped wind models are almost identical 
  (accounting for the different mass loss rates)
  and cannot be distinguished on this
  scale. The blue side of \ion{N}{5}\,$\lambda$1240 is affected
  by interstellar Ly\,$\alpha$ absorption. \label{W113Fig}}
\end{figure}

A straightforward consequence of the very good agreement between
the observed spectrum and the \tlusty\  model is that the wind
of MPG~113 is weak and the mass loss rate small. Indeed, no 
P\,Cygni line profiles are seen in the UV spectrum. The \ion{C}{4}
and the \ion{N}{5} resonance doublets show a weak signature of the
stellar wind (a weak, blueshifted, absorption components), while
\ion{O}{5}\,$\lambda$\,1371 does not reveal any indication of a
stellar wind. The comparison of MPG~368 and MPG~113 (Fig.~\ref{W368Fig}
and Fig.~\ref{W113Fig}) is quite striking: while the two stars have
a similar effective temperature, the spectral signatures of a wind
have almost completely vanished in MPG~113, most likely because of
its lower luminosity (about half the luminosity of MPG~368).
We shall return to this striking difference in Sect.~\ref{MlossDisc}
to examine the predictions of the radiatively-driven wind theory.

To derive the wind parameters of MPG~113, we have built a small grid of
CMFGEN models, starting with a mass loss rate derived from
\citet{vink01} formula and decreasing it until we match
the observed spectrum. We have adopted a standard wind acceleration,
$\beta = 1$, and the parameters derived from the photospheric analysis.
We can set an upper limit to the mass loss rate from the \ion{C}{4}
doublet, while an homogeneous wind model predicts a strong, unobserved,
\ion{N}{5} doublet, as well as an (unsaturated) P\,Cygni profile for
\ion{O}{5}\,$\lambda$\,1371. We examine the role of clumping too,
adjusting as before the mass loss rate in order to keep matching
the \ion{C}{4} lines. Similarly to the other stars, clumping improves
the fit of \ion{O}{5}\,$\lambda$\,1371, although not as dramatically.
The \ion{N}{5} lines are also predicted much weaker, but are still
predicted significantly too strong. Having set the abundances from
photospheric lines, we cannot achieve a good fit of \ion{C}{4},
\ion{N}{5}, and \ion{O}{5} simultaneously. We have adopted a mass
loss rate, \Mdot = 3\,10$^{-9}$\,\msolyr, for an homogeneous wind
(and a mass loss rate three times smaller for a clumped wind) as a
best compromise. The predicted \ion{C}{4} blue extended absorption is
too weak. The clumped model improves the fit to \ion{N}{5} which still
remains too strong and it provides a good match to \ion{O}{5}.
However, the only possible firm conclusion is an upper limit to
MPG~113 mass loss rate, \Mdot\ $< 10^{-8}$\,\msolyr. 

\subsection{NGC 346 MPG 487}
\label{MPG487Sect}

\begin{figure}
\figurenum{9}
\plotone{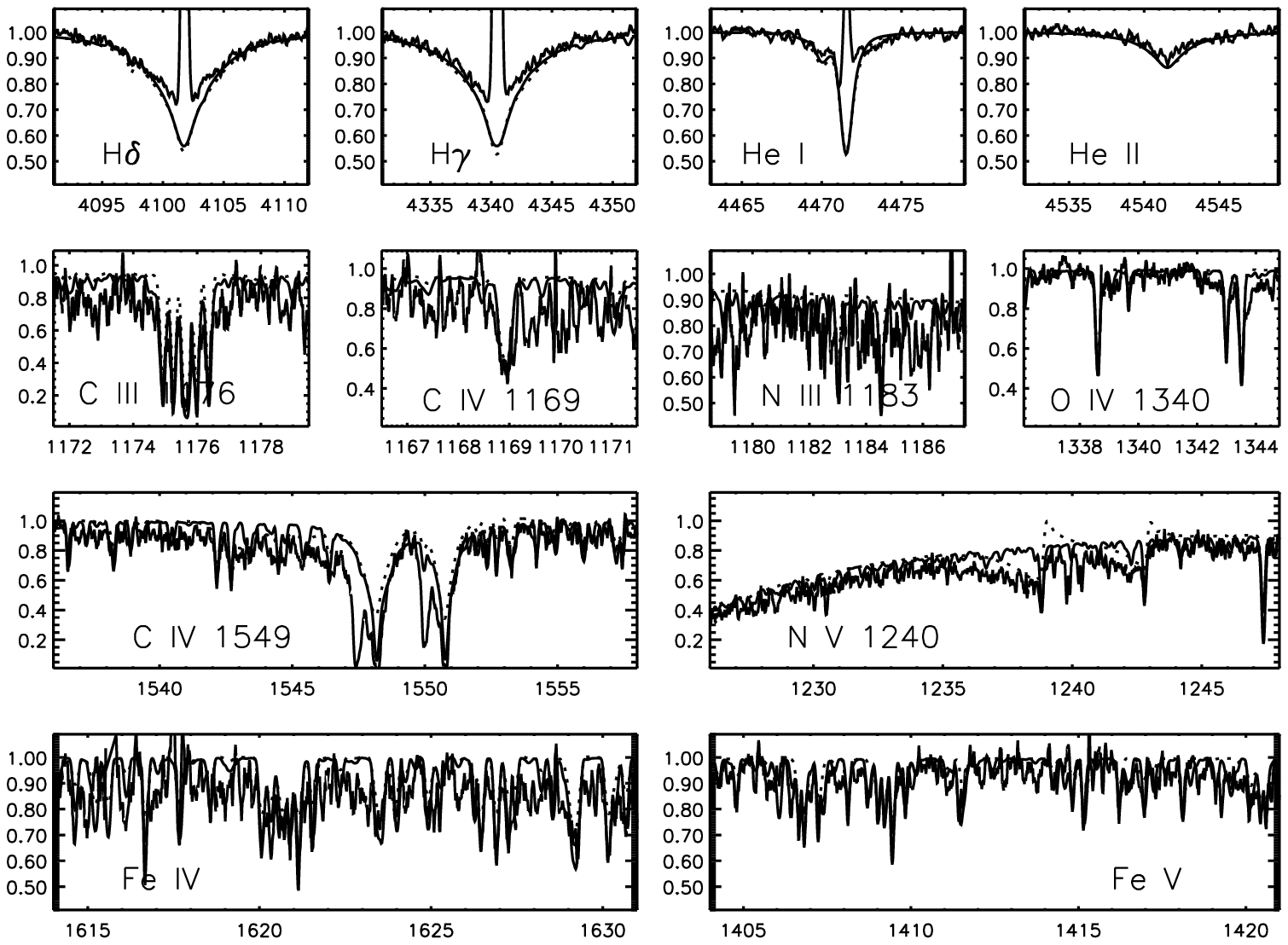}
\caption{Best fits to MPG 487 photospheric lines used to derive the stellar parameters,
  \teff=35000\,K, $\log g=4.0$, and surface chemical composition, $Z/Z_\odot=0.1$
   (full line: \tlusty, dotted line: CMFGEN). \label{P487Fig}}
\end{figure}

The hydrogen and helium optical lines as well as the UV \ion{C}{3}
and \ion{C}{4} lines can be matched with model atmospheres having
\teff\ = 35\,000\,K and $\log g$ = 4.0. The spectrum reveals two
main characteristics: {\it (i)} very narrow and weak metal lines;
{\it (ii)} very weak signatures of a stellar wind similarly to
MPG~113. Fitting the metal lines requires to adopt a low
microturbulent velocity, $\xi_{\rm t}$ = 2\,\kms, and a metallicity
of one tenth the solar value. A larger microturbulence, e.g. 5 or
10\,\kms, would imply an even lower metallicity. The \ion{S}{5}\,$\lambda$\,1502
is matched satisfactorily by a model spectrum with the lowest microturbulence. 
Having thus fixed the microturbulent velocity, we have to adopt an oxygen 
abundance, O/O$_\odot$=0.2, to match the \ion{O}{4}\,$\lambda$\,1338-43
lines. Once the effective temperature and the microturbulent velocity
have been fixed, we can derive the carbon abundance from
\ion{C}{3}\,$\lambda$\,1176, 1247 and \ion{C}{4}\,$\lambda$\,1169,
and the nitrogen abundance from \ion{N}{3}\,$\lambda$\,1182-84.
The carbon and nitrogen abundances, and thus the N/C abundance ratio,
indicates that the surface composition of MPG~487 is quite similar
to nebular material in the SMC. Fig.~\ref{P487Fig} displays the best
fits achieved with \tlusty\  and CMFGEN model atmospheres.

Despite the very low metallicity derived from the UV spectrum, we are
unable to match the visible spectrum. The observed metal lines are
very weak (e.g., \ion{Si}{4}\,$\lambda$\,4088-4116). Some predicted
lines (e.g., \ion{C}{3}\,$\lambda$\,4647-50, \ion{N}{3}\,$\lambda$\,4634-40)
are even not detected.
At face value, the optical spectrum would yield very low upper limits
(typically 0.01\,$\times$\,solar) for the abundances of these
elements with the adopted stellar parameters (35\,000; 4.0).

We suggest that our spectra are contaminated by the presence of
one or several additional  stars in the slit. As the simplest case,
let us imagine that two similar O stars are in the slit, the first
one with narrow lines and the second with very broad lines. This
would result in a diluted spectrum with weak narrow lines as observed
in the UV. The situation in the optical might even be worse due
to the fact that the atmospheric seeing could blend the light
of more contributors. New observations at very high spatial
resolution would therefore be extremely helpful to solve this
problem of very weak metal lines in this star. We caution therefore
about over-interpreting our results on MPG~487 at this stage.

Like MPG 113, this star reveals no conspicuous P~Cygni line profiles
in its UV spectrum. The only indication of mass loss is the
\ion{N}{5}\,$\lambda$\,1240 resonance doublet. No redshifted emission
component is observed, but absorption wings develop bluewards of the
rest wavelength of the doublet, extending up to $\approx -1100\pm 100$\,\kms.
Because of the low effective temperature that we derived from the photospheric
spectrum, \ion{N}{5} is not much populated. This detection might
suggest that the mass loss rate is not negligible, in order to produce the
observed blue absorption wings. However, the photospheric model spectrum
calculated from the \tlusty\  model reproduce already quite well
the \ion{C}{4}\,$\lambda$\,1549 resonance doublet, suggesting thus
that \Mdot\  should remain fairly small. Assuming the carbon and nitrogen
abundances determined from the other UV lines (see above), we have
calculated a series of CMFGEN models starting with the mass loss rate
derived with the \citet{vink01} formula (\Mdot\  $\approx 10^{-7}$\msolyr).
We have decreased the mass loss rate until reaching a satisfactory agreement
with the \ion{N}{5} and \ion{C}{4} doublets, yielding
\Mdot\  $\approx 3\,\times\,10^{-9}$\msolyr.
We stress however that this value remains uncertain because
\ion{N}{5} is a minor ion and that there is no real wind signature
in the \ion{C}{4} lines (see Fig.~\ref{P487Fig}).
An upper limit, \Mdot\  $< 10^{-8}$\msolyr, is nevertheless a robust result.
Due to the limited spectral information on the stellar wind, and in particular the
absence of \ion{O}{5}\,\lb1371 at this low temperature, we cannot establish any
meaningful constraint about clumping in the wind. Finally, we point out that
the very weak wind lines implies that all stars in the STIS slit must have
very weak winds. This might be the case if they all have a sufficiently low
luminosity (see \S\ref{EvolDisc}).

\subsection{NGC 346 MPG 12}
\label{MPG12Sect}

MPG~12 is located at the outskirts of the \ion{H}{2} region. Its spectrum is
much less contaminated by nebular emission allowing us to make good use of the
Balmer and the \ion{He}{1} lines in our analysis. MPG~12 was classified as a
late O-type star. This is the coolest star of our sample, as evidenced by the
weakness of \ion{He}{2} lines, of \ion{C}{4}\,\lb1169, and of \ion{Fe}{5} lines,
and the relative strength of \ion{He}{1} lines. We found a good match to the
Balmer, \ion{He}{1}, and \ion{He}{2} lines with (30\,000\,K; 3.5) and
(32\,500\,K; 3.75) \tlusty\  model atmospheres with a metallicity, $Z/Z_\odot=0.2$.
Although the fit seems slightly better for the first model, some degeneracy
in stellar parameter space does exist. We have adopted an intermediate
model, \teff\  = 31\,000\,K and $\log g=3.6$. This model also reproduces well
carbon and iron ionization balances as shown by UV \ion{C}{3-iv} and
\ion{Fe}{4-v} lines.

At this temperature, we expect  neither strong \ion{O}{4} lines nor 
\ion{S}{5}\,\lb1502. We cannot therefore set the microturbulent velocity in
the same way than we did with hotter stars. For this purpose, we used mainly 
iron lines. We found that a model assuming a microturbulent velocity,
$\xi_{\rm t}=5$\,\kms, together with a metallicity, $Z/Z_\odot=0.2$, reproduces
well the observed strength of iron lines. We have adopted these values.
An alternative solution would be a higher microturbulence (10\,\kms) together
with a lower overall metallicity (one tenth solar), but the adopted model provides
a better match to the observations (e.g., the silicon visible lines). The lower
metallicity ($Z/Z_\odot=0.1$) was favored earlier by \citet{AAS197}.
In the meantime, we have improved iron line collisional strengths and showed
that the earlier collisional data resulted in lower iron abundances up to
a factor 2 \citep{Tueb2}.

The carbon abundance was then determined in fitting \ion{C}{3}\,\lb\lb1176, 1247,
4647-51 and \ion{C}{4}\,\lb1169. We found that the nebular SMC abundance results
in good fits. Similarly, we could use several nitrogen lines, \ion{N}{3}\,\lb1183,
4097, 4634-40, and \ion{N}{4}\,\lb1718, to derive the nitrogen abundance. These
lines are matched with a solar nitrogen abundance. This results in an N/C abundance
ratio 30 times larger than the nebular ratio. This huge surface enrichment was already
noticed by \citet{walborn00} and \citet{AAS197}.

MPG~12 shows the weakest spectral wind signatures in our sample. Only slight
asymmetries can be observed in the \ion{C}{4} and possibly in the \ion{N}{5}
resonance lines. We calculated a series of CMFGEN models, proceeding similarly
to MPG~113 and MPG~487, starting from the mass loss rate predicted by
\citet{vink01} and decreasing the mass loss rate until \ion{C}{4}\,\lb1549
is predicted as weak as observed. Homogeneous wind models were assumed and we
kept $\beta$ fixed to unity. This required a quite low mass loss rate
for an O star, \Mdot\  $= 1\,\times\,10^{-10}$\,\msolyr. \citet{AAS197}
gave a higher mass loss rate (\Mdot\  $= 5\,\times\,10^{-10}$\,\msolyr) which
followed from their adopting a lower carbon abundance.
   
% ------------------------------------------------------------------------------

\section{Discussion}
\label{DiscSect}

Table~\ref{ResuTab} summarizes the results of our analysis. We conservatively
estimate uncertainties
to $\pm$5\% on \teff, $\pm$0.1\,dex on $\log g$, and $\pm$2 to $\pm$5\,\kms\ 
on $\xi_{\rm t}$ from low to high microturbulent velocities. We do not
find any evidence of helium enrichment, and we can thus set a limit on the
helium abundances, $y < 0.15$. Typical uncertainties on the other abundances
are $\pm$0.1 to $\pm$0.2\,dex. The mass loss rates of the three most luminous
stars are derived with an accuracy of 10-20\,\% from the fitting procedure (but this does
not include potential systematic errors), while we have mostly obtained limits
on the wind properties of the three other stars. We discuss now the implications
of our results, comparing them to other observational and theoretical studies.

\subsection{Effective Temperatures}
\label{TeffDisSect}

The \teff\ - Spectral Type relation for O stars used most commonly
during recent years for studies of young stars and young star-forming
regions with massive stars has been established by \citet{Vacca96}.
This relation is based on spectroscopic analyses performed in the late 1980's
and early 1990's with NLTE H-He model atmospheres. \citet{HHL98} 
pointed out that the inclusion of metal line blanketing in the NLTE
model atmospheres would result in deriving lower effective temperatures.
Indeed, several recent studies based on the newest generation of model
atmospheres that incorporates the effect of line blanketing point towards
lower \teff \citep{martins02, crowther02, herrero02, bianchi02}.

\citet{martins02} have established a new \teff\ -- Spectral Type calibration
for Galactic dwarf O stars. Their NLTE line-blanketed model atmospheres,
calculated with CMFGEN, predict that line blanketing will result in lowering
the effective temperature scale by $\Delta$\teff = 1500 to 4000\,K from late
to early O-type stars. They have also shown that $\Delta$\teff\  
is typically only 60\% at SMC metallicity (they adopted $Z/Z_\odot = 0.125$)
compared to changes derived at solar metallicity. This smaller change is
expected because the effect of line blanketing is reduced at lower metallicity.

We found effective temperatures that are about 7000\,K lower than 
\citet{Vacca96} relation for O4V stars, while they are  3000 to 3500\,K lower for
later O stars. Given the recent redefinition of the earliest spectral types
\citep{walborn02}, we cannot derive a meaningful conclusion
concerning MPG~355.  These changes in effective temperature are larger
than predicted by \citet{martins02}, especially for the two stars with early
spectral types. We point out however that the early spectral types attributed to
MPG~324 and MPG~368 may be an artefact due to the filling of \ion{He}{1} lines by
nebular emission. If these two stars were classified O6 instead, our results will
be more in line with Martins et~al. Fig.~\ref{TeffFig} displays
the \teff\ - Spectral Type relation for Galactic dwarf O stars from Vacca et~al.
and the improved relation from Martins et~al. This figure shows that our temperature
estimates are roughly similar or might be somewhat lower than Martin et~al.'s,
despite the lower metallicity
(and presumably a lesser blanketing effect). We have compared our \grid\ 
model atmospheres calculated at different metallicities, and we confirm that
only a part of the difference in \teff\  can be attributed to the omission
of line blanketing in earlier model atmospheres. Thus, we essentially agree with
Martins et~al. that the inclusion of line blanketing in model atmospheres lowers
the deduced effective temperatures, but in addition we argue that the effect is
larger than they predicted, likely because of the use of additional temperature
indicators like the carbon and iron ionization balance. The magnitude of
changes in \teff\  not as extreme as the changes found by \citet{crowther02}
for 4 O supergiants in the SMC and in the LMC. Further discussion
of the effective temperature scale of O stars at low metallicity will be presented
in a separate paper analyzing the full sample of 17 stars observed with HST/STIS
in the SMC \citep{heap03a}.
   
\subsection{Surface Abundances}
\label{AbDiscSect}

Our results on chemical abundances may be summarized in three points:
\begin{itemize}
\item[$\bullet$] These six stars do not reveal any helium enrichment at their surface;
\item[$\bullet$] however, carbon and nitrogren show non-solar abundance ratios;
\item[$\bullet$] the overall metallicity is $Z/Z_\odot=0.2$.
\end{itemize}
Analyzing a sample of Galactic O stars, \citet{herrero92}
found that most of them show a helium enrichment at their surface, which
was not expected from stellar evolution theory. They termed this problem
the ``helium discrepancy''. The origin of this discrepancy can be attributed
to their use of NLTE H-He model atmospheres for the spectroscopic analysis.
Stronger \ion{He}{1-ii} lines are predicted by NLTE line-blanketed model
atmospheres. Additionally,
\citet{smith98} argued that a significant microturbulent
velocity would strengthen the He lines. High helium abundances would thus
be spuriously derived if one assumes that NLTE model atmospheres of O stars
should not require any microturbulence to match stellar spectra.
Our analysis, based on NLTE line-blanketed model atmospheres and high-resolution
spectra, shows that a significant microturbulence is required to match the
UV metal lines, and yields solar-like helium abundances. While we cannot exclude
small helium enrichments, our results do agree with stellar evolution theory
predictions that the surface helium abundance remains mostly unchanged during the
main sequence phase of O stars. Even evolutionary models with rotationally-enhanced
mixing do not predict significant helium enrichment \citep{Grot7}.
We conclude therefore that the helium discrepancy problem most likely finds its roots
in neglecting the role of microturbulence and in neglecting metal line blanketing in
the model atmospheres.

In  order to interpret the CNO abundance patterns that we have determined in these
six stars, we need first to establish a basis for comparison, that is the composition
of the original material from which these stars were formed. \citet{venn99}
summarizes the results of abundances studies in the SMC. While evolved F-K SMC supergiants
yield C/N abundance ratios that are similar to the solar neighborhood and young Galactic
B stars (N/C$\approx$1/3), SMC nebular studies \citep{dufour84, garnett95}
provide a lower abundance ratio, N/C$\approx$1/6 \citep{venn99}. There are
still uncertainties on the nebular nitrogen abundance in NGC~346. We have
adopted a nebular nitrogen abundance, N/N$_\odot$=0.03, from \citet{dufour84}
and \citet{venn99}, which was supported by recent observations by
\citet{peimbert00}. However, with new photoionization models of NGC~346, Rela\~{n}o,
Peimbert, \& Beckman (2002) have argued that the nitrogen abundance might be
as high as N/N$_\odot$=0.08. They found however that the derived nebular nitrogen abundance
is very sensitive to the exact shape of the EUV stellar fluxes. In Table~\ref{ResuTab},
we have listed the derived N/C abundance ratios relative to the nebular ratio,
keeping the lower nebular nitrogen abundance. We thus point out that 
these relative N/C ratio may be reduced by a factor 2 if the
\citet{peimbert02} recent models turn out to predict the correct abundance.

All but one star (MPG~487 which has anomalously weak lines, see \S\ref{MPG487Sect}) show
a systematic surface enrichment of nitrogen. Moreover, all stars but MPG~355
have a carbon abundance that is consistent with the nebular abundance.
Canonical evolutionary models do not predict such a nitrogen surface enrichment
during the main sequence phase, but mixing
induced by fast rotation may result in such abundance pattern (e.g., Maeder~\&
Meynet 2001). Excluding the anomalous case of MPG~487, the OC6 star remains the
least chemically evolved (lowest N/C ratio), corroborating the finding that the
OC7.5 III((f)) star AV69 is less evolved than the O7 Iaf+ supergiant AV83 \citep{AV83}.
We further examine the evolutionary status of these
stars in the following section. \citet{heap03b} discuss the surface
abundances of the complete stellar sample. Oxygen cannot be derived from this dataset.
However, we can argue that our carbon abundances indirectly support the idea that
oxygen abundances have to be similar to the nebular value, O/O$_\odot$=0.15-0.19
\citep{venn99, peimbert00}. In this analysis, we have adopted
O/O$_\odot$=0.2, following the derived metallicity. Additional observations
of abundance-sensitive oxygen lines in the near-UV would therefore be a very valuable
complement to this study.

Lines of other elements (Si, S, Fe) are well reproduced by adopting scaled-solar
abundances. We have derived an overall metallicity, $Z/Z_\odot=0.2$, in good agreement
with previous studies, as summarized by \citet{venn99}. This value confirms
also previous analyses of O stars in the SMC \citep{haser98}.

\subsection{Evolutionary Status and Age}
\label{EvolDisc}

We compare in Fig.~\ref{HRFig} the location of the six studied stars in the HR diagram
with predictions of stellar evolution theory. We are using canonical tracks (``normal''
mass loss rates, no rotation) from the Geneva group with the appropriate metallicity,
$Z/Z_\odot=0.2$ \citep{Geneva3}. Three stars (MPG~324, 368, and 113)
have locations that coincide with an isochrone for $t\approx3\,10^6$\,yr. This agrees well
with an early estimate based on a NLTE analysis of optical spectra
\citep{kudr89}, while a younger age of 1\,Myr was recently given by
\citet{walborn00}. The later value is based on effective temperatures
derived from the spectral type, and the greater age derived from NGC~346 main sequence O stars
is now a direct consequence of our lower effective temperature scale.

MPG~355 supported the very young age of NGC 346 derived by Walborn et~al. A straightforward
interpretation of its location in the HR diagram indeed suggests a very young and
massive star ($M\approx 90\,M_\odot, t < 10^6$\,yr). Such a massive star would
have evolved off the main sequence after 3\,Myr. We argue however that so young a star could
not show already
such an enhancement of its nitrogen surface abundance. Its blue, luminous location
might rather be the consequence of an homogeneous evolution due to rotation near
break-up velocity \citep{maeder87b}. A 60\,$M_\odot$ star with a very fast
initial rotation ($v_{\rm rot}\approx 400$\,\kms) could have evolved towards MPG~355
location (see Fig.~8 in Meynet~\& Maeder 2000).
From our analysis, we can moreover exclude that MPG~324 and MPG~368 are as hot as
\teff$\geq$45\,000\,K. Such a high effective temperature is necessary to assign an age
as young as $t\approx 10^6$\,yr to these two stars. A young MPG~355 would therefore
not be coeval with them. The older age, $t\approx3\,10^6$\,yr, assuming a homogeneous
evolution for MPG~355, seems thus the better alternative.

The location of MPG 487 in the HR diagram indicates that this star is more evolved.
Yet, its very weak line
spectrum suggests contamination by one or several very close stars (see \S\ref{MPG487Sect}).
While the effective temperature is determined from line strength ratios, the total
luminosity would be overestimated. Therefore, the location of MPG~487 in the HRD
is a supportive hint of the multiple nature of the observed object. If
we attribute to MPG~487 a luminosity corresponding to the 3\,Myr isochrone, the
predicted mass loss rate would also be significantly lower and in better agreement
with our estimate (see \S\ref{MlossDisc}).

\citet{walborn00} already argued that MPG~12 is most likely not
a coeval member of the cluster, based on its spatial location outside the nebula,
its somewhat discrepant radial velocity, and the derived stellar parameters. Our
detailed analysis supports this conclusion, and we derive an age of 8\,Myr for this
star. MPG~12 is the most evolved star in our sample, showing the highest enhancement
in surface nitrogen as may be expected.

\subsection{Observed and Predicted Mass Loss Rates}
\label{MlossDisc}

We start by comparing our mass loss rates to \citet{puls96}
results for Galactic, LMC and SMC stars. Then, we examine
how our results compare to theoretical predictions
from radiative line-driven wind theory, with an emphasis on the
dependence of the mass loss rate on metallicity. First, we will stay
in the classical framework of homogeneous, smooth winds. We will then
examine the evidence for clumping in the winds of O stars, and
discuss the implications of our results for the wind momentum-luminosity
relation (WLR). Finally, we shall explore some alternative explanations
for the weak winds observed in some stars.

\subsubsection{Homogeneous Winds}

In the context of the CAK theory of radiatively driven winds 
\citep{cak75}, \citet{abbott82} calculated the line force and
discussed its dependence with metallicity. He suggested a dependence of
the mass loss rate in the form \Mdot\ $\propto Z^m$, with $m\approx 1$.
This metallicity dependence was reexamined in later works, resulting
in a weaker dependence: \citet{kudr87} obtained
$m\approx 0.5$, while \citet{Leit92} found
$m\approx 0.8$. The wind terminal velocities are also found to be
significantly smaller at lower metallicities.

The dependence of mass loss on metallicity is most likely the main influence
of metallicity on the evolution of massive stars. It is therefore
essential to specify better this dependence, in order to provide a
firmer basis to model massive stars in very young galaxies. We will first
explore this topic empirically, before comparing our results to new
theoretical predictions.

We plot in Fig.~\ref{MZFig} the mass loss rates derived for Galactic and SMC
stars. We use \citet{puls96} data, excluding supergiants,
and our results. To guide the eye, we have also plotted least-squares fits
for Galactic and SMC stars, excluding the three stars with very low mass loss
rates. As expected, the mass loss rate is smaller in SMC stars in comparison
to Galactic stars. Yet, this difference does not look so well established if we consider
only O dwarfs (filled symbols), as the three Galactic dwarf
stars are more luminous than their SMC counterparts.
The SMC relation is skewed by MPG~355 which is significantly more luminous than
other studied SMC stars and was included in the two studies. The least-squares
fits suggest that the mass loss rates are 0.2 to 0.6\,dex smaller in SMC stars
in comparison to Galactic values, with an increasing difference
for decreasing luminosities. At lower luminosities, three stars show a precipitous
drop of the mass loss rate. Assuming a metallicity, $Z/Z_\odot=0.2$,
the least-squares fit implies a dependence, \Mdot\ $\propto Z^m$,
with $m=0.6\,\pm\,0.3$.
While this empirical dependence appears in good agreement with theoretical
expectations, we point out that a small $m$ is mostly imposed by MPG~355.
Excluding this star, the power law factor is close to unity. A better definition
of the metallicity dependence will require additional {\em homogeneous}
analyses of Galactic and SMC main sequence stars.

The dependence of mass loss on stellar metallicity was recently
revisited theoretically by \citet{vink01}. We
compare now our results to their predictions.
Vink et~al. have calculated the mass loss rate for models covering a range
in effective temperature, luminosity, and metallicity. The major improvement
of their method over previous approaches is to take into account the
effect of multiple scattering via a Monte-Carlo technique when calculating
the transfer of momentum from the radiation field to matter. They found
a relation, \Mdot\ $\propto Z^{0.85}$. The relation is even flatter once
they factor in the effect of metallicity on the wind terminal velocity,
\Mdot\ $\propto Z^{0.69}$.

Table~\ref{MlossTab} lists the stellar parameters required to predict
the mass loss rates from Vink et~al.'s recipes (their Eq.~24). Since
the uncertainty on the stellar masses remain fairly large, we have adopted
typical masses for our objects based on the masses listed in Table~\ref{ResuTab}.
We have adopted a mass, $M=65\,M_\odot$, for MPG~355 based on our
argument that it might have undergone homogeneous evolution, and that
MPG~355 is rather unlikely to be more massive given NGC~346 age.
The escape velocities have been calculated using effective
masses derived with Eddington factors, $\Gamma_{\rm e} = 0.434, 0.206, 0.13$,
for $\log L = 6.0, 5.5, 5.0$ \citep{vink01}. The terminal
velocities of Galactic O star winds are related to the escape velocities as
$v_\infty\approx 2.6\,v_{\rm esc}$, with an accuracy of 20\%
\citep{groen89, lamers95, kudr00}.
The terminal velocities in SMC O stars are smaller by about 20\% \citep{Leit92},
and thus $v_\infty/v_{\rm esc}\approx 2.0$. We have measured terminal velocities
for the three most luminous stars which agree with these scalings. We could
only derive lower limits for the terminal velocities of the three other
stars, and we have therefore adopted terminal velocities that are twice the
escape velocities. We apply Vink et~al.'s Eq.~24 using a metallicity,
$Z/Z_\odot = 0.2$.

The predicted and measured mass loss rates for MPG~355, MPG~324 and
MPG~368, are in reasonable agreement. This conclusion holds however only
if we compare mass loss rates derived from homogeneous wind models
(as assumed by Vink et al.). While the agreement is excellent
for MPG~355, our mass loss rates measured for the two other stars are
about half the predicted ones. If we were to adopt a larger mass,
$M=90\,M_\odot$, for MPG~355, this conclusion would not be changed.
On the other hand, we find dramatic differences for the three other
stars: their mass loss rates are 1 to 2 orders of magnitude smaller
than predicted. This discrepancy may be smaller for MPG~487 if we
assume a lower luminosity to correct from its probable multiple nature.
Assuming $\log L\approx 4.6$ and $M\approx 20\,M_\odot$, corresponding
to a 3\,Myr old star with \teff =35\,000\,K, the predicted mass loss
rate is decreased to \Mdot =10$^{-8}$\,\msolyr. This is still a factor
3.5 higher than the observed value. We compare measured and predicted
mass loss rates in Fig.~\ref{MlossFig}, including also results from
\citet{puls96} and \citet{AV83}. While
Vink et~al.'s relation appears to provide a good agreement with mass loss
rates measured for evolved stars, it clearly overestimate the mass loss
rate of main-sequence SMC O stars. As an illustration, we display
in Fig.~\ref{Vink113Fig} the predicted profiles for MPG~113 calculated
assuming Vink et~al. mass loss rate. The mass loss rate is corrected
by a factor, $(f_\infty)^{0.5} = 0.1^{0.5}$, for the clumped model.
Clearly, we cannot match the observed line profiles with this
mass loss rate, or we would need very low, unrealistic CNO abundances.
\citet{AV83} found
mass loss rates that 2 to 3 times larger than the theoretical values. They
argued however that AV~83 mass loss is most likely enhanced due to fast rotation.
Fig.~\ref{MlossLFig} displays the ratio of predicted to measured mass loss
rates as a function of the stellar luminosity. Most clearly, the theoretical
predictions do not account for thin winds and low luminosity main sequence
stars ($L < 3\,10^5\,L_\odot$). Three brighter stars in Puls et~al. sample 
($L\approx 5\,10^5\,L_\odot$) also show much weaker winds than expected.
Furthermore, we point out that Vink et~al.'s predictions also overestimate
Puls et~al. measurements for giant/supergiants in the LMC. These results
may therefore suggest a steeper dependence than currently predicted
by the radiatively line-driven theory (\Mdot\  $\propto Z^{0.5-0.7}$).
Vink et~al.'s method consists in estimating the mass loss that can be
driven by the star to an observed $v_\infty$. Their method assumes but does
not show that such a mass loss can be driven through the sonic point.
This might be the underlying reason why they systematically overestimate
the mass loss rates.

Our understanding
of the mass loss dependence with metallicity remains therefore sketchy,
and cast some doubts on our ability to extrapolate the results obtained
on Magellanic Clouds stars to the very low metallicity environment
of young galaxies. Detailed studies of larger sample of O stars at
low metallicity would therefore be very valuable. We also need to
establish a larger reference set of Galactic stars using similar
high-quality spectroscopic data and the same modeling techniques.

\subsubsection{Clumped Winds}

Clumping in the wind introduces an additional complication in determining
the actual mass loss rates. In order to reproduce adequately the weak incoherent
electron scattering wings of \ion{He}{2} lines, \citet{hillier91} already
suggested over 10 years ago that the dense winds of Wolf-Rayet stars might
be clumped. Inclusion of clumping resulted in revising down by a factor 2
the mass loss rates of Wolf-Rayet stars \citep{hamann98}.
While \citet{puls96} have considered the effect on H$\alpha$
of clumping in O star winds, first evidences of clumping in the wind of O
supergiants were submitted only recently.
\citet{eversberg98} argued that stochastic variable substructures in \ion{He}{2}$\lambda$4686
reveal the presence of outmoving clumps in the wind of the O4 supergiant $\zeta$~Puppis.
\citet{crowther02} and \citet{AV83}  provided further arguments from spectroscopic analyses
that O supergiant winds are clumped, based namely on the \ion{P}{5} resonance
lines and on the absence of blue asymmetry in strong photospheric lines in AV~83.

The inability of standard, smooth O star wind models to reproduce
\ion{O}{5}\,\lb1371 is a long-standing problem. We point out that the
absorption and the emission components are always systematically predicted
too strong, by wind models assuming a $\beta$-velocity law as well as
by models solving for the hydrodynamical structure. This failure is therefore
not a consequence of adopting an empirical velocity/density law.
Several solutions have
been envisaged, including lower effective temperatures 
\citep{dekoter98} or lower oxygen abundances (e.g. Haser et~al. 1998).
While these two solutions decrease the \ion{O}{5} line strength, there
are no other supporting evidences for them; moreover, they
are not improving on the qualitative disagreement between predicted
and observed line shapes. In this paper, we have achieved for the first
time a good match of \ion{O}{5}\,\lb1371 line, assuming the simple
clumping recipes incorporated in CMFGEN. We used a volume clumping factor,
$f_\infty\approx 0.1$, consistent with previous studies of clumped winds,
for the three stars with \teff\,$\approx$\,40\,000\,K. A smaller clumping
factor, $f_\infty = 0.01$, yields a better fit for the hottest star, MPG~355,
but results in an unrealistic large line driving force. To drive MPG~355
observed wind, our calculations of line driving forces support clumping,
but with a filling factor somewhat larger than 0.1. At the current point,
we think that this shows that clumped models predict, correctly or by 
coincidence, the \ion{O}{4}/\ion{O}{5} density around the sonic point
and in the wind, and thus match \ion{O}{5}\,\lb1371. Lacking reliable estimates
of the oxygen abundance remains a major uncertainty. We feel therefore that
it is still somewhat premature to claim that O star winds are clumped.
However, since this conclusion has potentially foremost consequences for
stellar evolution in reducing the mass loss rates of massive stars, we believe
that clumping needs to be further investigated, building a consistent model
of \ion{O}{5}, \ion{P}{5}, and strong photospheric lines while deriving
oxygen abundances independently.

\subsubsection{Wind Asymmetries}

We have pointed out that our wind models predict too strong a
\ion{N}{5}\,\lb1240 emission component in all cases, even though we have not
accounted for the increased ionization due to shocks in the wind. We notice
that \citet{haser98} also predict too strong \ion{N}{5} emission
in their analysis of two SMC stars, MPG~355 and AV~243, although their
models fit the observed profile of two LMC stars well. Effective temperatures,
nitrogen abundances, and mass loss rates, have been derived by matching other lines.
Fitting the observed \ion{N}{5}\,\lb1240 line profiles would require incompatible
changes in these parameters. Clumping helps a little, but does not solve
the inconsistency fully. Adopting slightly lower nitrogen abundances might also
lessen this discrepancy.

On the other hand, we recall that our sample has been selected with a bias
towards stars with sharp-lined spectra. Given the fast intrinsic rotation of
O stars, we are viewing our sample stars preferentially pole-on.
\citet{owocki98} have argued that gravity darkening effects can lead to a
reduced mass loss, and thus lower density, in the equatorial region of the
wind, contrary to earlier studies of a Wind Compressed Disk model 
\citep{WCD93}. The lower equatorial density would thus result
in weaker emission components in stars seen pole-on. Further analyses of
unbiased stellar samples are necessary in order to support our suspicion.

\subsubsection{Wind Momentum-Luminosity Relation}

We display the modified Wind momentum-Luminosity relation (WLR) in
Fig.~\ref{WLRFig}, where we compare the WLR for Galactic and SMC stars. The wind
momentum of the three brightest stars of our sample (MPG~355, 324, and 368)
agrees well with values derived for other SMC stars. Least-squares fits
through Galactic supergiants and SMC stars ($\log L > 5.4$) have a very
similar slope, $1/\alpha^\prime\approx 1.9$, and are comparable to a value derived
by \citet{puls96}. The offset between the two relations
can be interpreted as an effect of metallicity and implies,
\Mdot $v_\infty \propto Z^m$, with $m = 0.87$. Allowing for a
metallicity dependence of the wind terminal velocities, $v_\infty \propto Z^{0.13}$
\citep{Leit92}, we obtain a dependence of the
mass loss rate, \Mdot\ $\propto Z^{0.74}$, which agrees well with theoretical
expectations. However, we should point out that the Galactic relation is established
for supergiants, while the SMC relation is based mostly on giants and
dwarfs. The few Galactic dwarfs fell clearly below the supergiant relation and,
therefore, the difference in wind momentum between Galactic and SMC dwarfs
is significantly reduced (by a factor 2 at least). On the other hand, evolved
SMC stars like AV~83 and AV~69 have a wind momentum that is quite similar to
these of Galactic supergiants.

A second interesting feature of the WLR is the steep turn at low luminosity
($\log L < 5.5$). This behavior was already noticed by
\citet{puls96} for a few Galactic and SMC stars, though the drop may start
at higher luminosity at low metallicity. The three low luminosity
stars in our sample confirm this steeper trend, as well as recent results
in the star forming region N81 in the SMC where low luminosity OVz stars
reveal very weak wind signatures \citep{martins03}.
This break in the WLR appears to be primarily related to the stellar luminosity,
not to the effective temperature
because low wind momentum spans a range in \teff, from 45\,000 to 30\,000\,K.
This break is therefore not related to bistability jumps that are predicted
by Vink et~al. (and references therein). They predict
a jump around 35\,000\,K with an increase of the mass loss rate toward lower
\teff, albeit at metallicities much lower (1/30 solar) than those considered here.

We have also plotted WLR data for three early B supergiants with
\teff\ $\approx$ 28\,000\,K \citep{kudr99, herrero02}.
The O supergiant WLR can be extended to the lower
luminosity, $\log L < 5$, for two of these B supergiants. This shows
therefore a striking difference in the wind properties of early B supergiants
and low luminosity O stars. A stronger wind might be driven by the
B supergiants due to their lower surface gravity, $\log g = 3.0-3.2$,
compared to O stars ($\log g = 4.0$).
Cooler B supergiants (\teff\,$<$\,25\,000\,K)
have a lower wind momentum because the wind is then driven by lines of
another set of ions \citep{kudr99}.
We are currently obtaining STIS spectra of B supergiants
in the SMC which will allow us to further study the WLR.

\subsubsection{Wind Decoupling}
\label{Decoupl}

Other mechanisms might account for the weak winds that we observe. First,
\citet{owocki99} suggested that for stars
with optically thin continuum near the sonic point, i.e stars
with low luminosities and/or low metallicity, 
the velocity field in the sonic region is such that the
line accelerations are significantly lower than previously considered 
in the Sobolev formalism. As a consequence, the wind properties are 
significantly modified, and mass-loss rates are smaller. 
Furthermore, investigating the instability 
caused by heavy (minor) ions runaway in low-density line-driven winds,
%Owocki \& Puls (\cite{owocki02}) have confirmed the findings of 
%Babel (\cite{babel95}, \cite{babel96}) and Krti$\check{c}$ka \& Kub\'at 
%(\cite{KK00}) about the decoupling of the metals ions with
%respect to the the bulk of the hydrogen-helium wind. They
\citet{owocki02} have 
derived a scaling relation for the stellar winds parameters at which
a decoupling between minor ions and the bulk of the hydrogen-helium wind
can occur. Applying this
relation to our sample, we found that decoupling may indeed occur
over most parts of the winds of MPG~12, 487, and 113.

\subsubsection{Hydrodynamics}

Because the CMFGEN models are fully line-blanketed, it is possible to
check the consistency between the derived wind parameters
(\Mdot, $v_\infty$, and $\beta$) and the computed line force.
For simplicity, we consider two stars: MPG~113 and MPG~355. To improve
the accuracy of the line force calculations, the models discussed here
include Ar, S, Cl, and Ni, in addition to the elements listed in Table~\ref{atom_mod}. 

We have derived a very low mass loss rate (\Mdot\ $< 10^{-8}$\,\msolyr) for MPG~113,
with inhomogeneous wind models having a mass loss rate a factor up to 10 lower.
With the derived mass-loss rate(s) the radiative acceleration is much larger than
that required to drive the wind. This is verified even for a model in which
we adopt, \Mdot\ = $10^{-8}$\,\msolyr, $f_\infty = 0.1$, and $v_\infty$ = 2200\,\kms.
In this model, the radiative acceleration is a factor of two too large in the outer region,
and exceeds that required to drive the wind except for two depths near the sonic
point. The discrepancy near the sonic point is not unusual -- the radiation force 
is very sensitive to the velocity law and to the adopted microturbulent velocity 
in this region (see a discussion in Hillier et~al. 2003).

How can we resolve this wind-momentum discrepancy? Three solutions readily
manifest themselves:
\begin{enumerate}
\item There is a problem in the computed CMFGEN ionization which affects the
       derived mass loss rates. This discrepancy could arise from several
       different sources including inaccurate atomic data, the 
       influence of X-rays, and CMFGEN's approximate treatment of clumping.
       We note (Sect.~\ref{MPG355Sect}) that many tests were performed to verify CMFGEN,
       and no major changes in the derived mass loss rate was obtained.
       Moreover, we note that the problem with fitting \ion{O}{5}$\lambda$1371 has
       been a problem in all other wind calculations (e.g., \citet{haser98}).
\item Wind decoupling (Sect.~\ref{Decoupl}).
\item The adopted $v_\infty$\  is too low. As noted, $v_\infty$\  cannot be derived from
       the observations. If $v_\infty$\  was much higher than 2000\,\kms, the
       discrepancy might be lowered.
\end{enumerate}

For MPG~355, $v_\infty$\  is well determined, and the principal uncertainty in checking 
the hydrodynamics is the volume filling factor. For $f_\infty = 1.0$, the radiation force is
too low to drive the adopted mass loss rate throughout the entire wind. 
The line radiation force needs to be increased by 20 to 90\% (depth-dependent, and
excluding the sonic region) to drive the wind. 

For $f_\infty = 0.1$, there is a better agreement with the hydrodynamics, in general.
In the outer region, the line force is too small by 20\%, while at $v_\infty/2$ it is
30\% too large. Given the uncertainties in the adopted velocity law and abundances, these
differences are within expected errors, and could probably be remedied with slightly different
but similar models. The line force is still much too small at the sonic point, but this
is not surprising given the large adopted turbulent velocity, $\xi_{\rm t}=$25\,\kms.

For $f_\infty = 0.01$, the line force is larger than that required to drive the wind
at all locations with $v(r) > $70\,\kms. Throughout the wind, the line force is too large
by at least a factor of two. We have not included Ar, S, Cl, and Ni, in this model; hence,
a full model will show an even larger discrepancy. It is thus difficult to see how
such a model could be adjusted to give consistency between both the observed
spectrum and the wind hydrodynamics.

% ------------------------------------------------------------------------------

\section{Conclusions}
\label{ConclSect}

We have analyzed the UV and optical spectrum of six young, massive O stars in
NGC~346, the largest \ion{H}{2} region in the SMC. We summarize here our main
findings:
\begin{itemize}
\item Similarly to several recent studies, we have derived lower effective temperatures
than values obtained from a standard relation with the spectral type. This results
in lower stellar luminosities and lower ionization fluxes from lower bolometric
corrections. Taking MPG~324 as an example, the effective temperature is lowered from
48\,670\,K (from O4V spectral type) to 41\,500\,K, decreasing the luminosity by a factor 1.6
and the number of Lyman continuum photons by a factor 2.
\item Some microturbulence is still required to match the UV metal lines. Our analysis hints
at a decrease of the microturbulent velocity from early to late O stars, from a value close
to the sonic velocity down to a low value, 2 to 5\,\kms. Having set the microturbulence independently
from UV metal lines, we are able to match the helium lines in the optical spectrum without
requiring a surface enrichment.
\item Most stars have a carbon abundance similar to the SMC nebular abundance, but
their surface is markedly enriched in nitrogen already during the MS phase. Either,
most are fast rotators viewed pole-on or mixing efficiency is underestimated by current
evolutionary models.
\item Young SMC O stars have an overall metallicity, $Z/Z_\odot=0.2$, that is consistent with
evolved F-K supergiants.
\item The dependence of the mass loss rate with metallicity is consistent with a relation,
\Mdot\ $\propto Z^m$, with $0.5 > m > 1.0$. A better characterization of $m$
requires larger stellar samples and detailed analysis of high-quality spectra. We have obtained
very low mass loss rates for three stars, which cannot be explain in the framework of
radiative line-driven wind, but may indicate decoupling between hydrogen, helium, and metals,
in the wind. \citet{vink01} predict significantly too high mass loss rates for
SMC and for LMC stars. Use or extrapolation of their relations at sub-solar metallicities
should be taken with caution. 
\item Homogeneous, smooth winds predict too strong \ion{N}{5}\,\lb1240 and \ion{O}{5}\,\lb1371
lines, suggesting that the winds of O dwarfs may be clumped or asymmetric, similarly
to winds of O supergiants and WR stars.
\item Finally, we have achieved a very good consistency between \tlusty\  and CMFGEN models.
This agreement demonstrates that we can use NLTE photospheric, static model atmospheres to derive stellar
parameters and abundances, even for the hottest O dwarfs with moderate winds (\Mdot\ $\approx$
few 10$^{-6}$\,\msolyr, assuming $\beta=1$).
\end{itemize}

\acknowledgments

This work was supported by a NASA-NRC Research Associateship at GSFC (JCB),
and by grants (GO 7437, AR 7985) from the Space Telescope Science Institute,
which is operated by the Association of Universities for Research in Astronomy, Inc.,
under NASA contract NAS5-26555. We thank the referee, Paul Crowther, for helpful
comments. TL enjoyed the warm mediterranean hospitality at
Laboratoire d'Astrophysique de Marseille during the completion of this paper.

% ------------------------------------------------------------------------------------

\clearpage

\clearpage

\clearpage

\clearpage

\clearpage

\clearpage

\clearpage

\clearpage

\clearpage

\clearpage

\begin{figure}
\figurenum{10}
\plotone{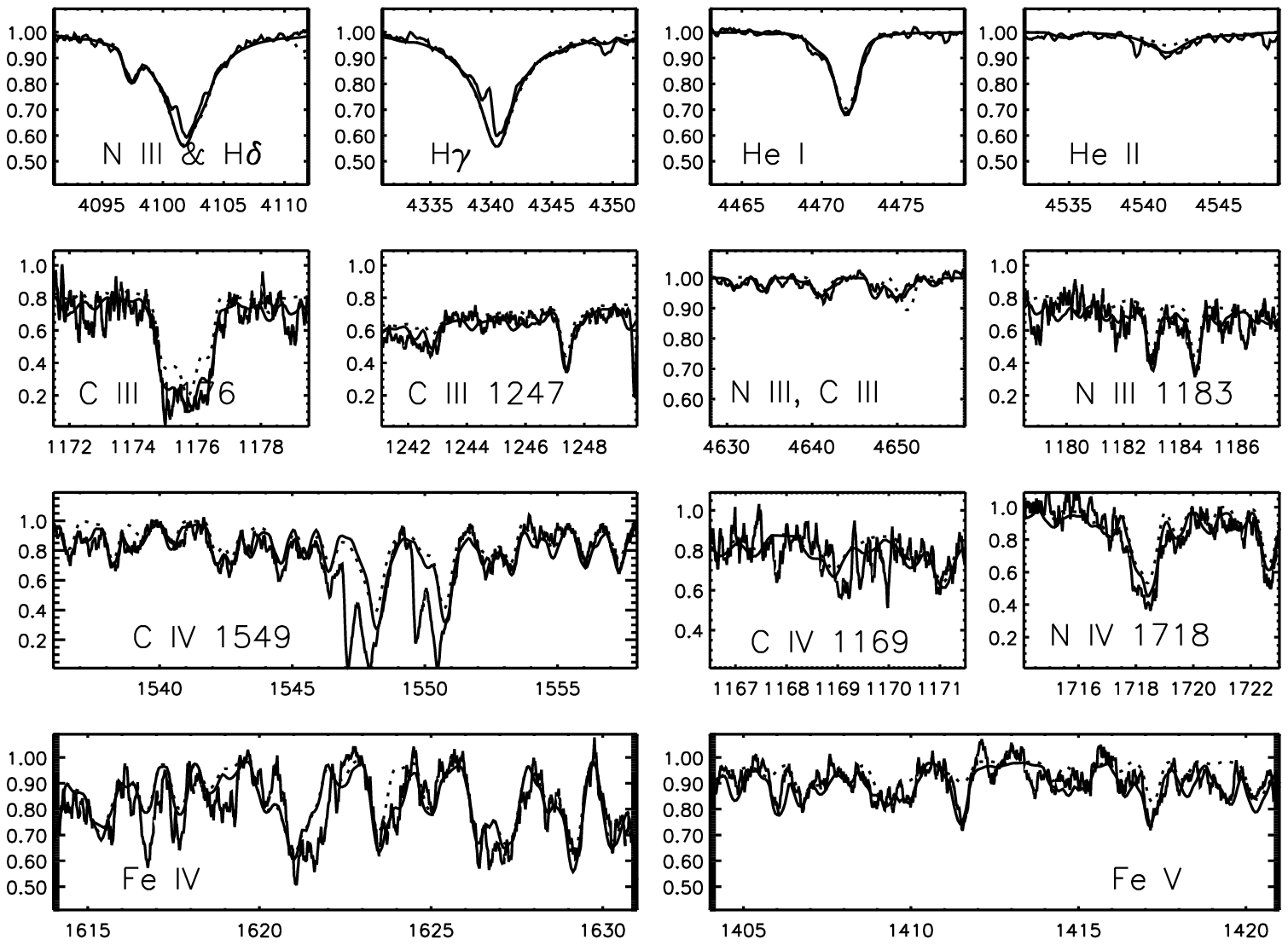}
\caption{Best fits to MPG 12 photospheric lines used to derive the stellar parameters,
  \teff=31000\,K, $\log g=3.6$, and surface chemical composition, $Z/Z_\odot=0.2$
   (full line: \tlusty, dotted line: CMFGEN). \label{P12Fig}}
\end{figure}

\clearpage

\begin{figure}
\figurenum{11}
\plotone{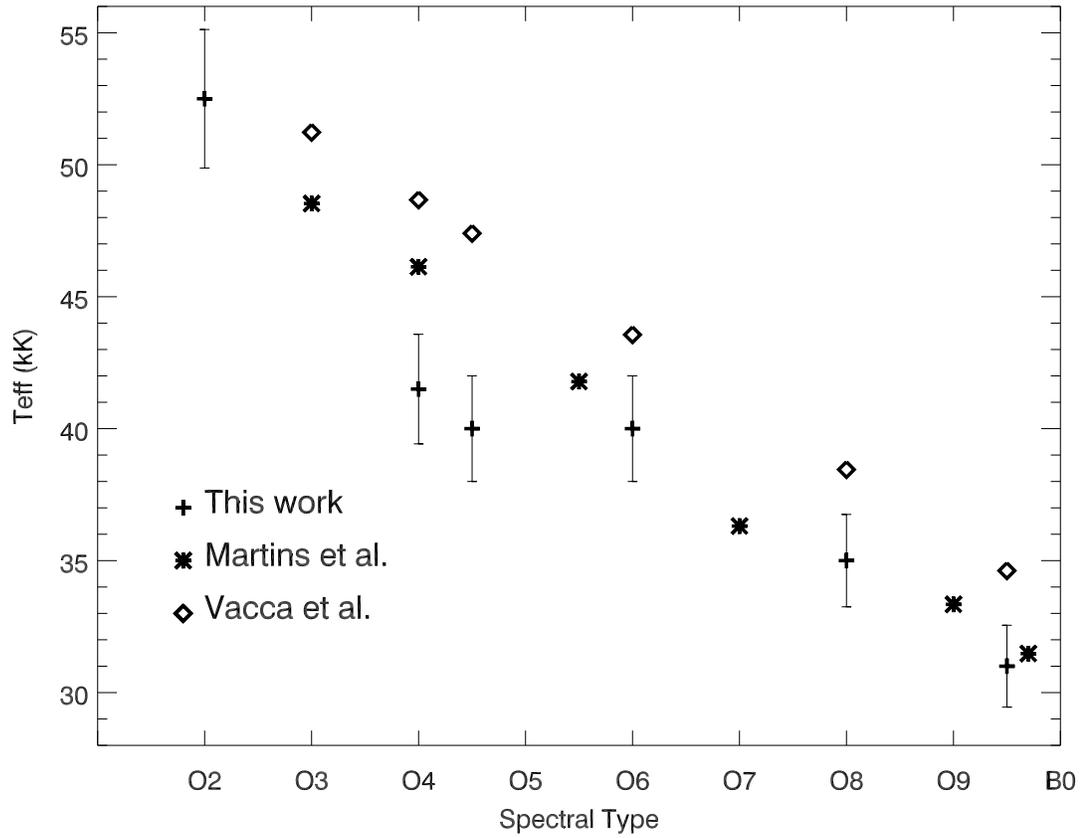}
\caption{\teff\ - Spectral Type relation for O dwarf stars
            from Vacca et al. (based on analyses of Galactic
            stars with unblanketed model atmospheres), from
            Martins et al. (based on blanketed CMFGEN model
            atmospheres with $Z=Z_\odot$), and from this study. \label{TeffFig}}
\end{figure}

\clearpage

\begin{figure}
\figurenum{12}
\plotone{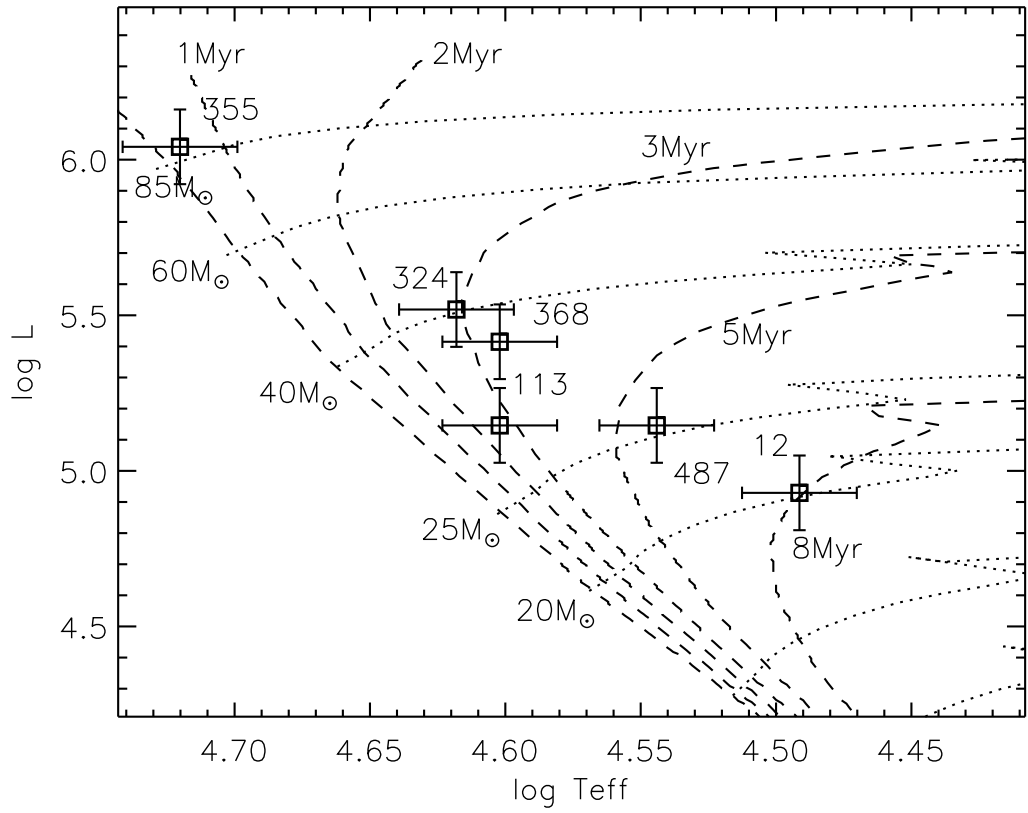}
\caption{HR diagram with Geneva evolutionary tracks for a SMC metallicity,
  $Z/Z_\odot=0.2$, and masses between 85 and 15\,$M_\odot$ (dotted lines), and
  isochrones (ZAMS, 1, 2, 3, 5 and 8\,$10^6$ yr; dashed lines). The large
  squares with error bars show our stellar sample in NGC 346. \label{HRFig}}
\end{figure}

\clearpage

\begin{figure}
\figurenum{13}
\plotone{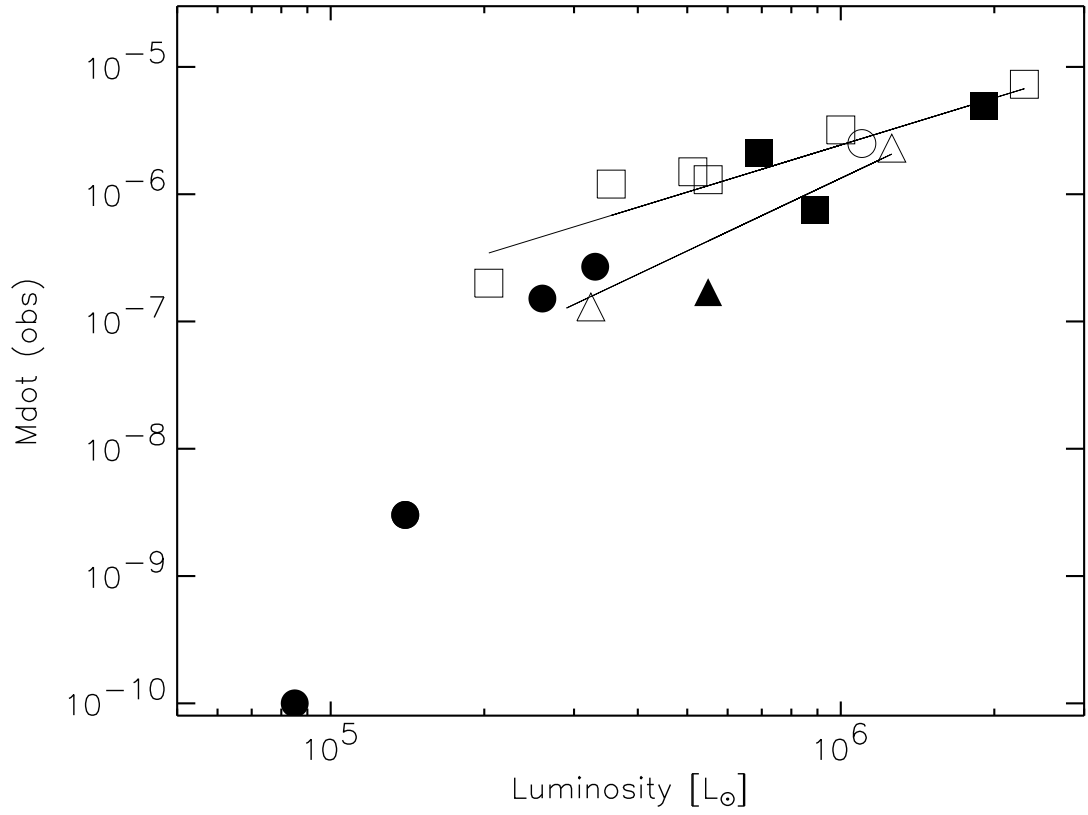}
\caption{Measured mass loss rates for Galactic and
  SMC O stars as a function of the stellar luminosity:
  squares: Galactic stars, \citet{puls96};
  circles: SMC stars, this paper; triangles: SMC stars, Puls et~al.;
  filled symbols: O dwarfs; empty symbols: O giants. Thin lines
  are least-squares fits for Galactic and SMC stars. \label{MZFig}}
\end{figure}

\clearpage

\begin{figure}
\figurenum{14}
\plotone{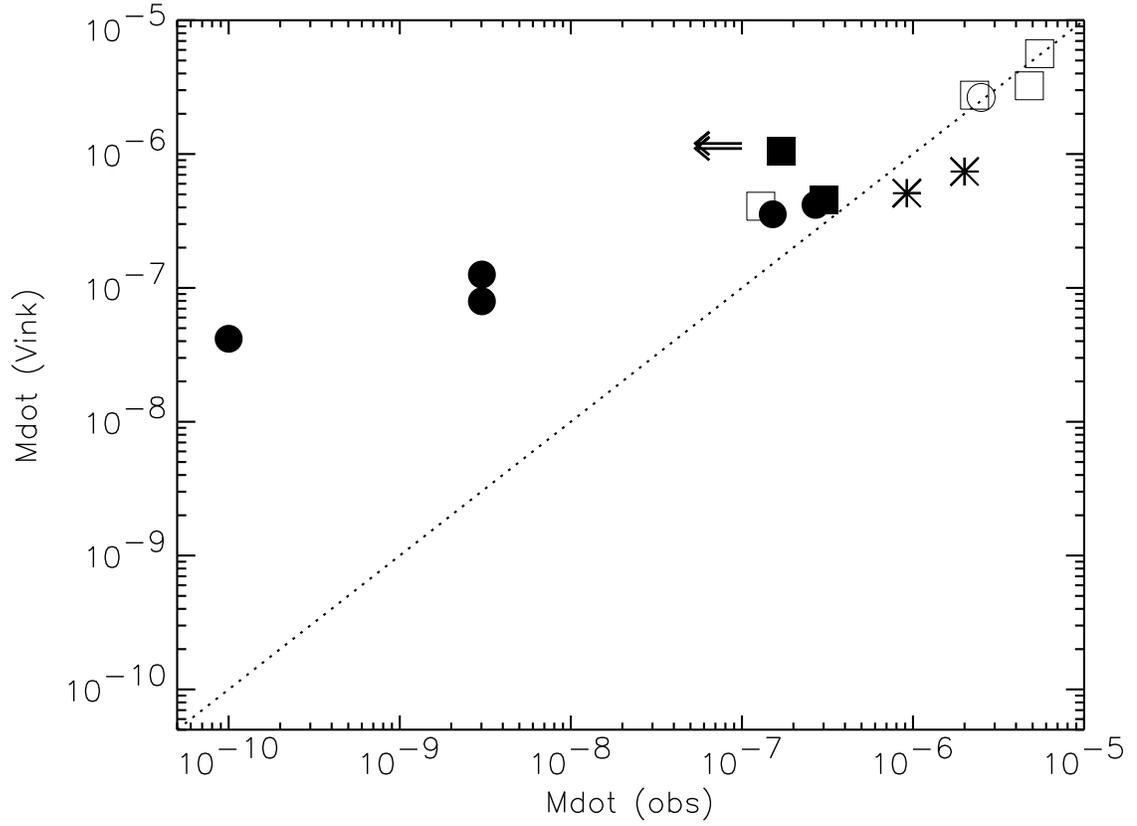}
\caption{Comparison of measured and predicted mass loss rates for
  SMC O stars: circles: this paper; squares: \citet{puls96};
  stars: \citet{AV83}; filled symbols: O dwarfs;
  empty symbols: giant/supergiant O stars. The double arrow indicates an
  upper limit. \label{MlossFig}}
\end{figure}

\clearpage

\begin{figure}
\figurenum{15}
\plotone{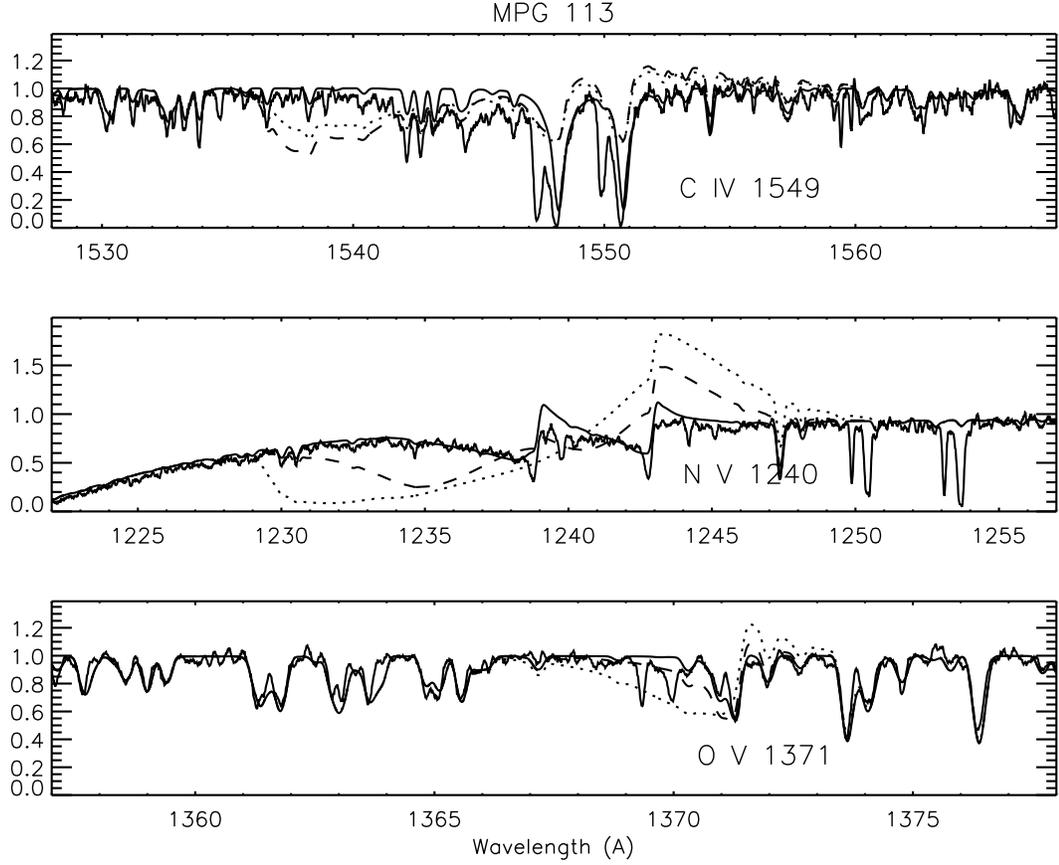}
\caption{Predicted wind profiles for MPG~113, for an homogeneous ($f_\infty=1.0$, 
  dotted line) and a clumped wind ($f_\infty=0.1$, dashed line),
  assuming Vink et al. theoretical mass loss rate (\Mdot\ = $10^{-7}$\,\msolyr)
  and predicted terminal velocity, $v_\infty=$2225\,\kms. Because of the weak
  P~Cygni profiles, $v_\infty$ cannot be easily deduced from the observations.
  The best model fit (clumped wind, $f_\infty=0.1$, \Mdot\ = $10^{-9}$\,\msolyr) is
  also shown for comparison (full line). \label{Vink113Fig}}
\end{figure}

\clearpage

\begin{figure}
\figurenum{16}
\plotone{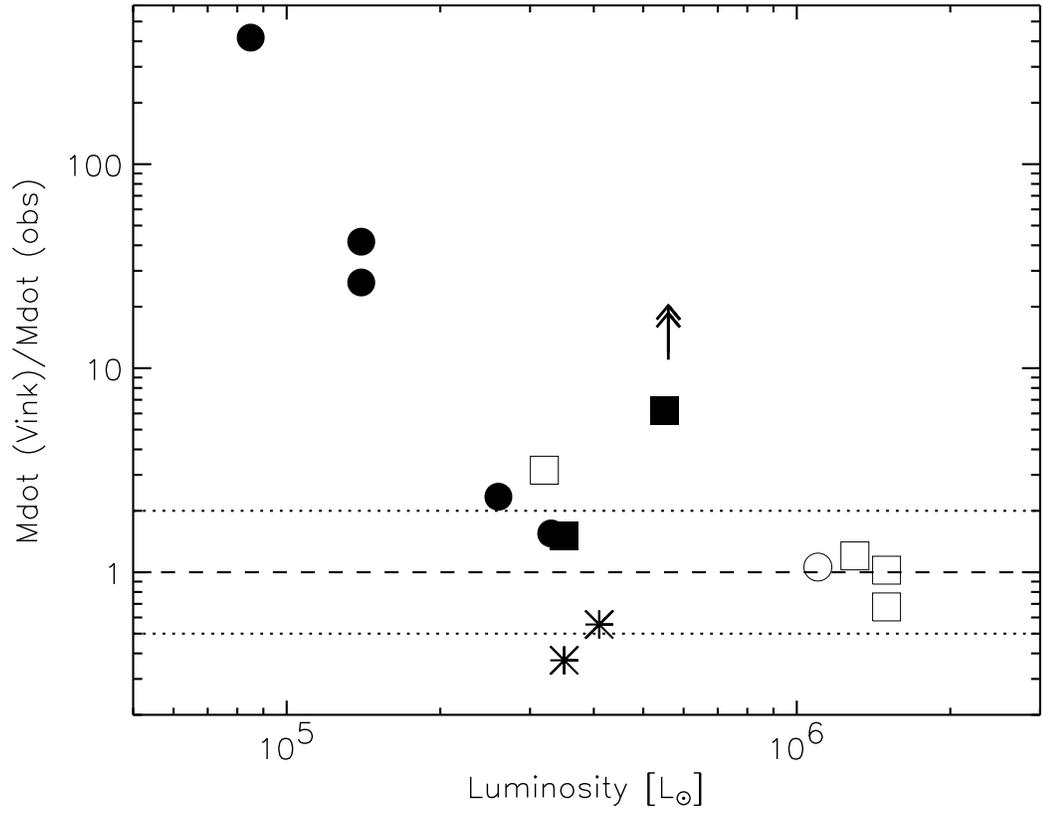}
\caption{Ratio of predicted to observed mass loss rates for SMC stars
     as a function of the stellar luminosity. Symbols as in Fig.~\ref{MlossFig}.
     The dotted lines indicate an agreement within a factor 2.\label{MlossLFig}}
\end{figure}

\clearpage

\begin{figure}
\figurenum{17}
\plotone{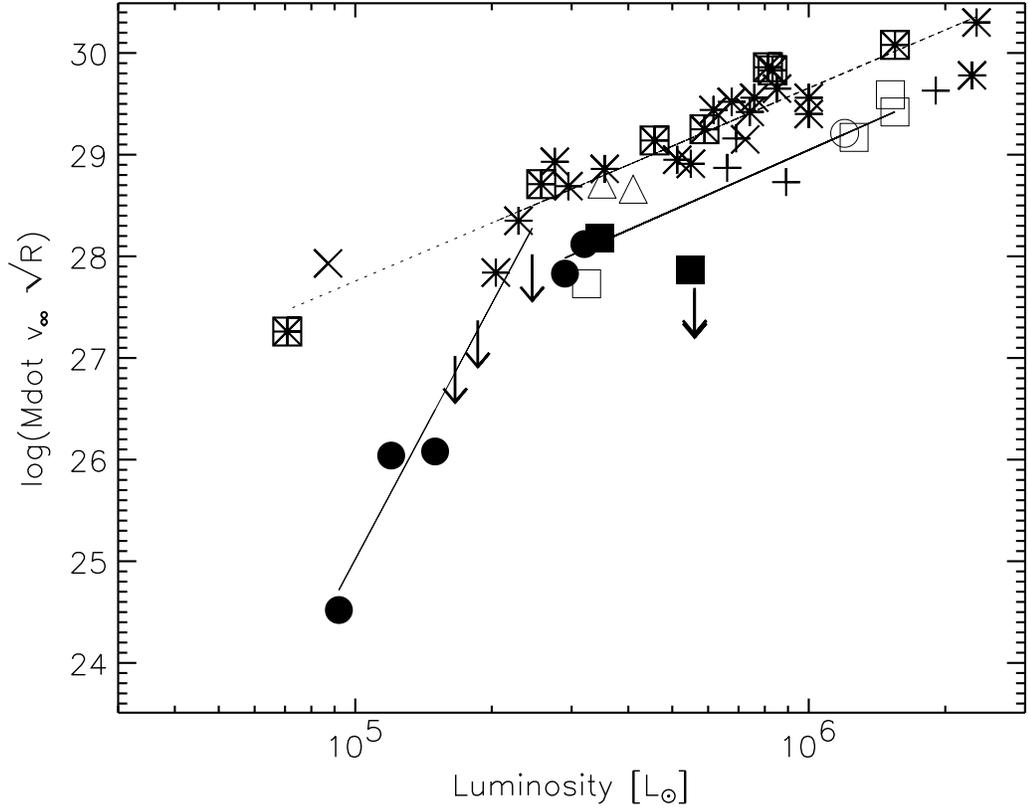}
\caption{Modified Wind momentum-Luminosity relation for SMC and Galactic stars.
     SMC stars: this work (circles), Puls et~al. (squares), Hillier et~al. (triangles);
     filled symbols: O dwarfs, open symbols: O giants/supergiants. Galactic
     supergiants: Puls et~al. (stars), Herrero et~al. (stars within boxes), Kudritzki et~al.
     (crosses); Galactic
     O dwarfs: Puls et~al. (plusses). Arrows represent upper limits.
     Least-squares fits are shown for Galactic supergiants (top dotted line),
     SMC stars (lower full line), and low luminosity stars (steep line). \label{WLRFig}}
\end{figure}

\clearpage

\begin{deluxetable}{llccccc}
\tabletypesize{\small}
\tablecaption{Observational characteristics of the sample stars. Radial velocities have been
          measured from the optical and UV spectra. Color excesses have been derived
          from intrinsic colors of \grid\  model atmospheres. \label{obsstis}}
\tablehead{
\colhead{Star} & \colhead{Spectral Type}   & \colhead{$ V$}  & \colhead{$B-V$}  &
 \colhead{$E(B-V)$}  &\colhead{$V_{r}$ [\kms]} & \colhead{Telescope}       }
\startdata
 NGC 346 MPG 355   & O2 III (f$^{*}$)  & 13.50 & -0.23  & 0.06   & 165     & AAT \\
 NGC 346 MPG 324   & O4 V((f))         & 14.02 & -0.24  & 0.05   & 200     & AAT \\
 NGC 346 MPG 368   & O4-5 V((f))       & 14.18 & -0.23  & 0.05   & 150     & AAT \\
 NGC 346 MPG 113   & OC6 Vz            & 14.93 & -0.22  & 0.07   & 180     & ESO \\
 NGC 346 MPG 487   & O8 V              & 14.53 & -0.22  & 0.07   & 160     & AAT \\
 NGC 346 MPG 12    & O9.5-B0 V (N str) & 14.87 & -0.15  & 0.12   & 220     & ESO \\
\enddata
\end{deluxetable}

\clearpage

\begin{deluxetable}{llll}
\tabletypesize{\small}
\tablecaption{Model atoms included in \tlusty\  and CMFGEN models. \label{atom_mod}}
\tablehead{
\colhead{Element} & \colhead{Ions}   & \colhead{\tlusty}  & \colhead{CMFGEN}      }
\startdata
    &                   & Superlevels       & \\
    &&& \\
H   & I, II             & 9, 1              & 20, 1 \\
He  & I, II, III        & 24, 20, 1         & 27, 22, 1 \\
C   & II, III, IV, V    & 22, 23, 25, 1     & 14, 30, 33, 1 \\
N   & II, III, IV, V, VI& 26, 32, 23, 16, 1 & 0, 34, 44, 41, 1 \\
O   & II, III, IV, V, VI, VII & 29, 29, 39, 40, 20, 1 & 24, 24, 29, 41, 13, 1  \\
Ne  & II, III, IV, V    & 15, 14, 12, 1     & 14, 23, 17, 1 \\
Si  & III, IV, V        & 30, 23, 1         & 20, 22, 1   \\
P   & IV, V, VI         & 14, 17, 1         & 36, 16, 1  \\
S   & III, IV, V, VI, VII   & 20, 15, 12, 16, 1 & 13, 51, 31, 28, 1 \\
Fe  & III, IV, V, VI, VII, VIII  & 50, 43, 42, 32, 1, 0 & 0, 100, 61, 57, 14, 1 \\
Ni  & III, IV, V, VI, VII  & 36, 38, 48, 42, 1 & \\
    &&& \\
 & & Individual Levels & \\
    &&& \\
H   & I, II             & 80, 1             & 30, 1 \\
He  & I, II, III        & 72, 20, 1         & 27, 30, 1 \\
C   & II, III, IV, V    & 44, 55, 55, 1     & 14, 54, 38, 1 \\
N   & II, III, IV, V, VI& 71, 68, 58, 55, 1 & 0, 34, 70, 49, 1 \\
O   & II, III, IV, V, VI, VII & 62, 116, 94, 89, 55, 1  & 25, 45, 48, 78, 13, 1 \\
Ne  & II, III, IV, V    & 29, 20, 18, 1     & 48, 71, 52, 1    \\
Si  & III, IV, V        & 105, 53, 1        & 34, 33, 1  \\
P   & IV, V, VI         & 14, 43, 1         & 178, 62, 1  \\
S   & III, IV, V, VI, VII   & 27, 15, 12, 34, 1 &  28, 142, 98, 58, 1 \\
Fe  & III, IV, V, VI, VII, VIII  & 12\,660, 13\,705, 11\,986, 4\,740, 1, 0 & 0, 1000, 300, 439, 153, 1 \\
Ni  & III, IV, V, VI, VII  & 11\,335, 13\,172, 13\,184, 13\,705, 1 &  \\
\enddata
\end{deluxetable}

\clearpage

\begin{deluxetable}{lllllll}
\tabletypesize{\small}
\tablecaption{Stellar parameters and chemical abundances derived for six O stars
          in NGC~346. \label{ResuTab}}
\tablehead{
\colhead{Star} & \colhead{MPG 355}   & \colhead{MPG 324}  & \colhead{MPG 368}  & 
                 \colhead{MPG 113}   & \colhead{MPG 487}  & \colhead{MPG 12}    }
\startdata
Spectral Type       & O2 III(f*)          & O4 V((f))           & O4-5 V((f))         & OC6 Vz              & O8 V                & O9.5-B0 V (N str)   \\
\teff\  [K]         & 52500               & 41500               & 40000               & 40000               & 35000               & 31000       \\
$\log g$ (cgs)      & 4.0                 & 4.0                 & 3.75                & 4.0                 & 4.0                 & 3.6        \\
$L$ [$L_\odot$]     & 1.1\,$\times\,10^6$ & 3.3\,$\times\,10^5$ & 2.6\,$\times\,10^5$ & 1.4\,$\times\,10^5$ & 1.4\,$\times\,10^5$ & 8.5\,$\times\,10^4$ \\
$\xi_{\rm t}$ [\kms]& 25                  & 15                  & 15                  & 10                  & 2                   & 5                 \\
$V \sin i$ [\kms]   & 110                 & 70                  & 60                  & 35                  & 20                  & 60               \\ [2mm]
Homogeneous winds   &                     &                     &                     &                     &                     &                  \\
\Mdot\ [\msolyr]    & 2.5\,\evi           & 2.7\,\evii          & 1.5\,\evii          & 3\,\eix         & 3\,\eix              & 1\,$\times\,10^{-10}$   \\
\vinf\ [\kms]       & 2800                & 2300                & 2100                & $\geq$1250          & $\geq$1100          & $\geq$1000     \\
$\beta$             & 0.8                 & 1.0                 & 1.0                 & 1.0                 & 1.0                 & 1.0             \\ [2mm]
Clumped winds   &                     &                     &                     &                     &                     &                  \\
$f_\infty$          & 0.01                &  0.1                & 0.05               &  0.1              &                  \\
\Mdot\ [\msolyr]    & 1.8\,\evii           & 1.0\,\evii         & 5.7\,\eviii        &  1\,\eix            &                    &    \\ [2mm]
Abundances\tablenotemark{a}  &                     &                     &                     &                     &                     &                  \\
$y$ (He/H)          & 0.1                 &  0.1                &  0.1                & 0.1                 &  0.1                &  0.1             \\
C/C$_{\odot}$       & 0.2                 &  0.06               &  0.06               & 0.1                 &  0.1                &  0.06  \\
N/N$_{\odot}$       & 1.0                 &  0.2                &  0.6                & 0.2                 &  0.03               &  1.0  \\
O/O$_{\odot}$       & 0.2                 &  0.2                &  0.2                & 0.2                 &  0.2                &  0.2  \\
Si/Si$_{\odot}$     & 0.2                 &  0.2                &  0.2                & 0.2                 &  0.1                &  0.2  \\
S/S$_{\odot}$       & 0.2                 &  0.2                &  0.2                & 0.2                 &  0.1                &  0.2  \\
Fe/Fe$_{\odot}$     & 0.2                 &  0.2                &  0.2                & 0.2                 &  0.1                &  0.2  \\ [2mm]
Evolutionary status\tablenotemark{b} &                     &                     &                     &                     &                     &                  \\
Age [$10^6$\,yr]    & $<$1 ?               & $\approx$3          &  $\approx$3         & $\approx$3          &  $\approx$5 ?        &  $\approx$8 \\
$M_{\rm evol}$ [$M_\odot$] &   90 ?         &  40                 &  38                 &  33                 &  25 ?                &  21 \\
$M_{\rm spec}$ [$M_\odot$] &   64         &  44                 &  26                 &  24                 &  32 ?                &  16  \\
N/C\tablenotemark{c} &  10                 &  6                  &  20                 &  4                  &  0.6                &  30 \\
\enddata
\tablenotetext{a}{Solar abundances from \citet{Sun98}.}
\tablenotetext{b}{Question marks indicate doubtful values, see further discussion in \S\ref{EvolDisc}.}
\tablenotetext{c}{Abundance ratio relative to the SMC nebular ratio, N/C$\approx$0.16 \citep{venn99}.}
\end{deluxetable}

\clearpage

\begin{deluxetable}{lllcrcrrr}
\tabletypesize{\small}
\tablecaption{Stellar parameters and predicted mass loss rates from 
            \citet{vink01}. Terminal velocities in parentheses
            are calculated as twice the escape velocities. \label{MlossTab}}
\tablehead{
\colhead{Star} & \colhead{\teff}   & \colhead{$\log L$}  & \colhead{$M$}  & 
                 \colhead{$v_{\rm esc}$}   & \colhead{$v_\infty$}  & \colhead{$v_\infty/v_{\rm esc}$} &
                 \colhead{log \Mdot (obs)} & \colhead{log \Mdot (Vink)} \\
               & \colhead{[K]} & \colhead{[$L_\odot$]} &\colhead{[$M_\odot$]} & \colhead{[\kms]} &
                 \colhead{[\kms]} & & \colhead{[\msolyr]}& \colhead{[\msolyr]}  }
\startdata
MPG 355  & 52500  &   6.04   &  65  &  1033         &  2800      &   2.71                 &  $-$5.6         &  $-$5.6  \\
MPG 324  & 41500  &   5.51   &  40  &  1053         &  2300      &   2.18                 &  $-$6.6         &  $-$6.4  \\
MPG 368  & 40000  &   5.41   &  40  &  1041         &  2100      &   2.02                 &  $-$6.8         &  $-$6.4  \\
MPG 113  & 40000  &   5.13   &  30  &  1112         & (2225)     &   2.00                 &  $-$8.5         &  $-$6.9  \\
MPG 487  & 35000  &   5.13   &  25  &   939         & (1880)     &   2.00                 &  $-$8.5         &  $-$7.1  \\
MPG 12   & 31000  &   4.93   &  20  &   795         & (1590)     &   2.00                 &  $-$10.0         &  $-$7.4  \\
\enddata
\end{deluxetable}

\end{document}